\documentclass[aps,prb,twocolumn,superscriptaddress,floatfix]{revtex4-1}
\usepackage{amsmath}
\usepackage{amssymb}
\usepackage{amsthm}
\usepackage{amsfonts}
\usepackage{graphicx}
\usepackage[usenames,dvipsnames]{xcolor}
\usepackage{tikz}
\usepackage{pgffor}
\usepackage{verbatim}
\usepackage{bbm}
\usepackage{float}
\usepackage{mathrsfs}
\usepackage{bbm}
\usepackage{multibib}
\usepackage{enumerate}
\usepackage[colorlinks, breaklinks=true,linkcolor=red, citecolor=blue, linktocpage=true]{hyperref}

\newcommand{\<}{\langle}
\renewcommand{\>}{\rangle}
\newcommand{\be}{\begin{equation} }
\newcommand{\ee}{\end{equation} }
\newcommand{\ba}{\begin{eqnarray} }
\newcommand{\ea}{\end{eqnarray} }

\newcommand{\mac}{\mathcal}

\newcommand{\bpm}{\begin{pmatrix}}
\newcommand{\epm}{\end{pmatrix}}
\newcommand{\bmm}{\begin{matrix}}
\newcommand{\emm}{\end{matrix}}

\newcommand{\bea}{\begin{eqnarray}}
\newcommand{\eea}{\end{eqnarray}}

\newcommand{\beq}{\begin{equation} }
\newcommand{\eeq}{\end{equation} }

\newcommand{\sig}{\text{sig}}
\newtheorem{lemma}{Lemma}

\newcommand{\mz}{\mathbb{Z}}

\newcommand{\bit}{\begin{itemize}}
\newcommand{\eit}{\end{itemize}}

\begin{document}

\title{Criteria for protected edge modes with $\mz_2$ symmetry}

\author{Chris Heinrich}
\author{Michael Levin}
\affiliation{Department of Physics, James Frank Institute, University of Chicago, Chicago, Illinois 60637,  USA}

\begin{abstract}
We derive a necessary and sufficient criterion for when a two dimensional gapped many-body system with Abelian anyons and a unitary $\mz_2$ symmetry has a protected gapless edge mode. Our criterion is phrased in terms of edge theories --- or more specifically, chiral boson edge theories with $\mz_2$ symmetry --- and it applies to any bosonic or fermionic system whose boundary can be described by such an edge theory. At an operational level, our criterion takes as input a chiral boson edge theory with $\mz_2$ symmetry, and then produces as output a prediction as to whether this edge theory can be gapped without breaking the symmetry. 
Like previous work, much of our derivation involves constructing explicit perturbations that gap chiral boson edge theories. Interestingly, however, we find that the standard class of gapping perturbations --- namely cosine terms constructed from null-vectors --- is not sufficient to gap some edge theories with $\mz_2$ symmetry, and thus we are forced to go beyond the usual null-vector analysis to establish our results.
\end{abstract}

\maketitle

\tableofcontents

\section{Introduction}

Some two dimensional gapped quantum many-body systems have the property that their edge is guaranteed to be gapless as long as certain fundamental symmetries are not broken. Systems of this kind are said to have \emph{protected} edge modes. A famous example is the 2D topological insulator, whose edge is gapless as long as time reversal and charge conservation symmetry are preserved.\cite{HasanKane, QiZhang}

An important aspect of protected edge modes is that they are a property of an entire gapped quantum \emph{phase}, not just a particular model. That is, if a many-body system has protected edge modes, then this property is shared by every other many-body system that can be continuously connected to the system in question without breaking a symmetry or closing the bulk energy gap. 

The physics of protected edge modes has been investigated most systematically for two classes of gapped phases. The first class consists of phases that have anyon excitations but no fundamental symmetries. Phases of this kind are often called `topological' phases. The second class consists of phases that have some set of symmetries but do not support anyon excitations. Phases of this type are known as `symmetry-protected topological' (SPT) phases. 

In both of these cases, there is a simple criterion that determines whether or not a given phase has a protected edge mode. For example, in the case of Abelian topological phases, it has been argued\cite{levin2013protected, KapustinSaulina} that the edge can be gapped (i.e. the edge is \emph{un}protected) if and only if (1) the chiral central charge $c_- = 0$, and (2) there exists a collection of anyons that forms a `Lagrangian subgroup.'\footnote{See section \ref{anyon_review} for the definition of a Lagrangian subgroup.} The criterion for non-Abelian topological phases is similar, but with Lagrangian subgroup replaced with a more complicated mathematical object, called a `Lagrangian algebra.'\cite{fuchs2013, konganyoncondensation2014} Likewise, in the case of SPT phases, it is believed that the edge can be gapped if and only if the SPT phase is `trivial.'\cite{ChenGuLiuWen} Here an SPT phase is said to be trivial if it can be realized by a completely uncorrelated many-body state like a product state or an atomic insulator.

Much less is known about protected edge modes in phases with \emph{both} global symmetries and anyons --- so-called `symmetry-enriched topological' (SET) phases. Most results have been derived only for individual phases;\cite{lu2012theory, HungWan2013, LuVishwanathz2sl, Burnelldoublesemion2016} general criteria for protected edge modes are lacking except for special symmetry groups.\cite{LevinStern2009,Neupertetal,levin2012classification}

To be clear, the problem is not just finding a criterion for protected edge modes, but finding one that is \emph{practical}. For example, it is possible to write down a general criterion based on the `gauging' construction, but this criterion is difficult or impossible to use in most cases. The criterion we have in mind\cite{HeinrichLevinunpub} applies to any SET phase that is built out of bosons and has a finite unitary onsite symmetry group $G$.\footnote{This criterion can be derived from the corresponding criterion for topological phases \emph{without} symmetry, using the correspondence between gapped boundaries of SET phases and gapped boundaries of gauged SET phases.\cite{HeinrichLevinunpub}} It states that SET phases of this kind can have a gapped symmetric edge if and only if three conditions are satisfied: (1) $c_- = 0$, (2) there exists a collection of anyons in the corresponding `gauged' SET phase that form a Lagrangian algebra, and (3) this Lagrangian algebra contains at least one anyon excitation carrying gauge flux $C_g$ for each conjugacy class $C_g \subset G$. (Here, the gauged SET phase is defined by coupling the original SET phase to a dynamical gauge field with gauge group $G$\cite{levin2012braiding}). To see the problem with this criterion, consider the simplest class of SET phases: those with Abelian anyons and an Abelian symmetry group $G$. Even in this case, the criterion becomes unwieldy if the symmetries permute different anyon species. The reason is that the gauged SET phase is \emph{non-Abelian} in this case\cite{BarkeshliWen, BarkeshliWenzn} and it is not clear how to find all Lagrangian algebras in a non-Abelian topological phase.

Thus, for all practical purposes we still do not have a systematic understanding of which SET phases have protected edge modes. The goal of this paper is to address this problem in the simplest case: SET phases with Abelian anyons and with a unitary $\mz_2$ (Ising) symmetry. Our main result is a necessary and sufficient criterion for when these phases can have a gapped symmetric edge. 

In general, SET phases can be described from either a bulk\cite{BarkeshliBondersonChengWang,TeoHughesFradkin,TarantinoLindnerFidkowski} or edge\cite{levin2012classification, HungWan2013, LuVishwanathz2sl, QiJianWang} point of view. In this paper we describe SET phases using edge theories, or more specifically, chiral boson edge theories with $\mz_2$ symmetry. Our criterion takes such a chiral boson edge theory as input and then predicts whether or not it can be gapped without breaking the symmetry. 
One limitation of this approach is that we can only analyze phases whose boundaries can be described by chiral boson edge theories. We do not know if every Abelian SET phase with $\mz_2$ symmetry has this property.

Our results are broadly similar to those of Ref.~\onlinecite{levin2012classification}, which derived an analogous criterion for systems with charge conservation and time reversal symmetry. However, there are several important differences between this work and Ref.~\onlinecite{levin2012classification}, beyond the choice of symmetry groups. First, while Ref.~\onlinecite{levin2012classification} established the necessity of their criterion using flux insertion arguments that are specific to $U(1)$ symmetry, we use a more general argument that may be applicable to \emph{any} symmetry group. In particular, we show that every gappable SET edge theory can be associated with a corresponding SPT edge theory which is also gappable, and then we establish the necessity our criterion by bootstrapping from the SPT case. 

The other difference between the two works involves how we show that our criterion is \emph{sufficient} for having a gapped edge. In this work, as in Ref.~\onlinecite{levin2012classification}, we establish the sufficiency of our criterion by constructing explicit perturbations that gap chiral boson edge theories. However, unlike Ref.~\onlinecite{levin2012classification} and many previous studies of SET edge theories,\cite{LevinStern2009, Neupertetal, lu2012theory, HungWan2013, LuVishwanathz2sl} we find that the standard class of gapping perturbations --- namely cosine terms constructed from `null-vectors' --- are \emph{not} sufficient to gap some edge theories (e.g. see appendix \ref{0101}), and thus we are forced to go beyond the usual null-vector analysis to establish our results. 

This paper is organized as follows. In section \ref{review_sect}, we review relevant formalism, including chiral boson edge theories, gapping perturbations, and Lagrangian subgroups. In section \ref{main_result}, we present our criterion and illustrate it with several examples. In sections \ref{boseSPTder} and \ref{fermSPTder}, we derive our criterion in the special case of bosonic and fermionic SPT edge theories, and in section \ref{SETder} we derive it in the general case. Technical arguments are given in the appendices.


\section{Review}\label{review_sect}

\subsection{Chiral boson edge theories}\label{review1}

Chiral boson theories are a general class of gapless one dimensional field theories which can be realized at the edges of two dimensional gapped many-body systems with Abelian anyons.\cite{Wenedgereview, frohlichedge, Wenbook} These field theories are built out of $N$ fields $\Phi_1,...,\Phi_N$, obeying commutation relations of the form
\beq
[\Phi_i(x'),\partial_x\Phi_j(x)]=2\pi iK^{-1}_{ij}\delta(x-x'),
\label{comm_rel_phi}
\eeq
together with a Hamiltonian of the form
\beq
H = \sum_{i,j=1}^N \int_{-\infty}^{\infty}  \frac{V_{ij}}{4\pi} \partial_x\Phi_i\partial_x\Phi_j dx
\eeq
Here $K$ is a $N\times N$ symmetric, non-degenerate integer matrix while $V$ is a $N\times N$ symmetric, positive definite real matrix. One can think of $K$ as describing the topological properties of the edge, while $V$ contains non-universal information, such as the velocities of the edge modes. In this paper, we will mostly ignore $V$ since our focus is on topological properties of edge theories.

Chiral boson edge theories come in two types: those for which $K$ has only even elements on the diagonal, and those for which $K$ has at least one odd element on the diagonal. We will refer to the first type of edge theory as `bosonic' and the second type as `fermionic.' The reason for this terminology is that the first type of edge theory describes the boundary of a many-body system built out of bosons, while the second type describes the boundary of a many-body system built out of fermions. 

An important ingredient in any edge theory is the set of local operators. For the above edge theories, the fundamental local operators are of the form 
$\mathcal O = e^{i \Lambda^T \Theta(x)}$ where 
\begin{align}
\Theta = \bpm \Theta_1 \\ \vdots \\ \Theta_N \epm, \quad \quad \Theta_j \equiv \sum_{k=1}^N K_{jk}\Phi_k
\end{align}
and $\Lambda$ is an $N$ component integer column vector. All other local operators can be constructed by taking derivatives and products of these operators. 

The physical interpretation of $e^{i\Theta_j(x)}$ is that it is an `electron' creation operator for the $j$th edge mode. Likewise, $e^{i \Lambda^T \Theta(x)}$ can be interpreted as a generalized electron scattering operator --- a product of electron creation and annihilation operators on different edge modes. (Note that in this paper we use the term `electron' to refer to the underlying microscopic particles from which the system is built, whether or not they are actually electrons).

\subsection{Including a $\mz_2$ symmetry}
To describe a $\mz_2$ symmetry transformation $S$ within a chiral boson edge theory, it suffices to specify the action of $S$ on the `electron' creation operators $e^{i \Theta_j}$. In this paper, we will consider symmetry actions of the following general form:
\begin{align}\label{z2symmetry_2}
S^{-1} e^{i\Theta_j} S =  e^{ i\sum_k W_{kj}\Theta_k} \cdot e^{i\pi\chi_j}
\end{align}
Here $W$ is an $N \times N$ integer matrix and $\chi$ is an $N$ component real vector. The matrix $W$ can be thought of as describing how the electron operators are permuted or mixed by the $\mz_2$ symmetry, while $\chi$ describes the additional phases that are accumulated. 


As an aside, we should mention that while it is tempting to express the symmetry transformation (\ref{z2symmetry_2}) as
\begin{align*}
S^{-1} \Theta_j S = \sum_{k=1}^N W_{kj} \Theta_k + \pi \chi_j,
\end{align*}
(as is common in the literature\cite{levin2012classification, lu2012theory, LuVishwanathz2sl, HungWan2013}) the latter formulation is less general {and we will not use it here}. The problem is that there is no reason that a symmetry action on $e^{i \Theta_j}$ must be extendable to a symmetry action on its non-compact counterpart $\Theta_j$. In fact, in many examples such an extension does not exist.\footnote{For example, consider the edge theory (\ref{K22_3}): the commutation relation $[\Theta_1(x), \Theta_2(x')] = \pi i [\text{sgn}(x'-x) + 1]$ (\ref{commklein}) is consistent with the symmetry transformation $S^{-1} e^{i \Theta_1} S = e^{i \Theta_2}$ but is inconsistent with $S^{-1} \Theta_{1} S = \Theta_2$. Interestingly, examples like this do not exist for \emph{anti-unitary} $\mz_2$ symmetries like time reversal, so in this case the non-compact formulation is equally general.}

Returning to Eq.~(\ref{z2symmetry_2}), we should point out that the matrix $W$ cannot be chosen arbitrarily: consistency requires that $W$ satisfy certain constraints:
\beq\label{symmetry_constraints_1}
W^2=1, \quad W^T K W=K
\eeq
Here the first constraint comes from demanding that $S^2 = 1$, while the second constraint comes from taking the logarithmic derivative of Eq.~(\ref{z2symmetry_2}) with respect to $x$, and then
imposing the fact that the symmetry must preserve the commutation relations (\ref{comm_rel_phi}). Note that the relation $S^2=1$ also gives constraints on $\chi$, but we will not need them here.

To summarize, a chiral boson edge theory with $\mz_2$ symmetry is characterized by a triplet $(K,W,\chi)$ obeying certain constraints (\ref{symmetry_constraints_1}). The matrix $K$ describes the commutation relations of the underlying fields and $W, \chi$ describe how the $\mz_2$ symmetry transformation $S$ acts on these fields.

\subsection{Gapping perturbations and null vectors}\label{gappertsect}
One of the goals of this paper is to determine which of the above edge theories can be gapped without breaking the $\mz_2$ symmetry and which cannot. To investigate this question, it is useful to have a library of perturbations that can be used to gap chiral boson edge theories. In this paper, we will focus on gapping perturbations that take the form
\begin{equation}
H' = \sum_{i=1}^{N/2} \int_{-\infty}^{\infty} U\cos(\Lambda_i^T\Theta -\alpha_i) dx
\label{perturbations}
\end{equation}
where $\Lambda_1,...,\Lambda_{N/2}$ are linearly independent integer vectors and $\alpha_1,...,\alpha_{N/2}$ are arbitrary phases. Here each cosine term can be thought of as describing a process where electrons are scattered between different edge modes. The reason we consider a sum of $N/2$ cosine terms is that this is the minimum number needed to gap an edge with $N$ modes.

Conveniently, there is a well-known sufficient\footnote{In fact, this condition is both necessary and sufficient.\cite{levin2013protected}} condition under which the above perturbation $H'$ is guaranteed to gap the edge if $U$ is large. In particular, it is known that the perturbation $H'$ will gap the edge if the vectors $\{\Lambda_1,...,\Lambda_{N/2}\}$ satisfy\cite{Haldane1995stability}  
\beq\label{null_criterion}
\Lambda_i^T K \Lambda_j=0\ \ \text{for all $i$ and $j$}
\eeq
Such vectors are commonly referred to as `null-vectors.' 

While the above null-vector condition (\ref{null_criterion}) guarantees that the perturbation $H'$ will gap the edge, we need to impose two more conditions to ensure that the gapped edge does not break the $\mz_2$ symmetry $S$. The first condition is that 
\begin{equation}\label{symmetry_criterion}
S^{-1} H' S = H'. 
\end{equation}
This condition guarantees that the perturbation does not break the symmetry explicitly. The second condition is that the set of null vectors $\{\Lambda_1,...,\Lambda_{N/2}\}$ is \emph{primitive}, that is, there is no solution to the equation
\beq
a_1\Lambda_1+\ldots +a_{N/2}\Lambda_{N/2}=k\Lambda
\label{primcond}
\eeq
where $a_i$ are integers with no common divisors, $k$ is an integer that is greater than $1$, and $\Lambda$ is an integer vector. This primitivity condition ensures that the perturbation does not break the symmetry \emph{spontaneously}.\cite{levin2012classification,weaksymmbreakingWangLevin}

If a perturbation $H'$ satisfying the above conditions exists, then it follows that the edge can be gapped without breaking the $\mz_2$ symmetry. However, it is important to keep in mind that the converse statement is not true in general, i.e. the absence of an appropriate set of null vectors does not imply that the edge cannot be gapped. We will see an example of this below when we consider the edge theory (\ref{K22_3}).

%
\subsection{Anyons and Lagrangian subgroups} \label{anyon_review}

An important aspect of chiral boson edge theories is that they contain information about the anyon excitations in the 2D \emph{bulk}. In particular, given an $N$-component edge theory with matrix $K$, the bulk anyon excitations can be parameterized by $N$-component integer vectors $l$, where the exchange statistics $\theta_a$ of an anyon $l_a$ and the mutual statistics $\theta_{ab}$ between two anyons $l_a, l_b$ are given by\cite{Wenedgereview, Wenbook} 
\begin{equation}
\theta_a = \pi l_a^T K^{-1} l_a, \quad \theta_{ab}=2\pi l_a^TK^{-1}l_b. 
\end{equation}
If $l$ is of the form $l = K \Lambda$ where $\Lambda$ is an integer vector, then the corresponding anyon is \emph{electron-like} --- that is, built out of electrons. Likewise, if $l- l' = K \Lambda$, then $l$ and $l'$ are said to correspond to \emph{equivalent} anyons, since they differ by the addition of some number of electrons. 

 
Why are bulk anyon excitations relevant to edge theories? One reason is that the properties of these anyons determine whether or not the chiral boson edge theory can be gapped. In particular, it is known that, in the absence of any symmetry, a chiral boson edge theory can be gapped if and only if (i) $\sig(K) = 0$ and (ii) the set of anyons contains a \emph{Lagrangian subgroup}.\cite{levin2013protected, KapustinSaulina} Here the definition of a Lagrangian subgroup depends on whether the system in question is bosonic or fermionic. In the bosonic case, a Lagrangian subgroup is defined as a subset of anyons $\mathcal{L}$ such that\cite{KapustinSaulina} 
\begin{itemize}
\item All the anyons in $\mathcal{L}$ have trivial mutual statistics with one another. 
\item All the anyons in $\mathcal{L}$ have bosonic exchange statistics.
\item Every anyon  not in $\mathcal{L}$ has nontrivial mutual statistics with at least one anyon in $\mathcal{L}$. 
\end{itemize}
Equivalently, a Lagrangian subgroup is a set of integer vectors $\mathcal{L}$ such that
\begin{itemize}
\item $l^TK^{-1}l'$ is an integer for any $l, l'\in\mathcal{L}$.
\item $l^TK^{-1}l$ is an \emph{even} integer for any $l \in\mathcal{L}$.
\item If $l'$ is not equivalent to any element of $\mathcal{L}$, then $l^TK^{-1}l'$ is non-integer for some $l \in\mathcal{L}$.
\end{itemize}
In the fermionic case, a Lagrangian subgroup is defined in the same way, except that we do not require that all the anyons in $\mathcal{L}$ have bosonic exchange statistics. Equivalently, we do not require that $l^TK^{-1}l$ is an \emph{even} integer for every $l \in\mathcal{L}$. 

For our purposes, it is also important to define the notion of when a Lagrangian subgroup $\mathcal{L}$ is \emph{invariant} under the $\mz_2$ symmetry: we will say that $\mathcal{L}$ is invariant under a $\mz_2$ transformation $S$ if the symmetry partner of each anyon in $\mathcal{L}$ is also in $\mathcal{L}$. In terms of the above parameterization, this corresponds to the condition that for each $l\in\mathcal{L}$ the vector $W^Tl$ is equivalent to some element $l' \in \mathcal{L}$.


\section{Main result}\label{main_result}
In this section we present our main result: a necessary and sufficient criterion for when a $\mz_2$ symmetric chiral boson edge theory $(K,W,\chi)$ can be gapped without breaking the symmetry. Before stating the criterion, we first define some important auxiliary quantities, which we denote by $\chi_+$ and $K_\pm$.

\subsection{The auxiliary vector $\chi_+$}
Given a chiral boson edge theory with $\mz_2$ symmetry, $(K,W,\chi)$, we define $\chi_+$ as any $N$ component real vector that satisfies the two conditions
\begin{align}
S^{-1} e^{i\Lambda_+^T\Theta} S &= e^{i\Lambda_+^T\Theta} \cdot e^{i\pi\Lambda_+^T\chi_+}, \nonumber \\
W^T \chi_+ &=\chi_+, 
\label{implicit_definition}
\end{align}
where the first equality holds for all integer vectors $\Lambda_+$ with $W \Lambda_+=\Lambda_+$.
The physical meaning of $\chi_+$ is similar to $\chi$ in that it describes the extra phases that are acquired by electron operators under the $\mz_2$ symmetry $S$. The main difference is that $\chi_+$ only describes the symmetry transformation properties of a \emph{subset} of electron operators, namely those that transform into themselves multiplied by a phase. 


{A note of caution about the above definition: to compute $\chi_+$ for a given edge theory, one needs to know how operators of the form $e^{i \Lambda_+^T \Theta}$ transform under the symmetry. These symmetry transformations can be derived from the general rule (\ref{z2symmetry_2}), but one has to be careful to take into account the non-trivial commutation relations for the $e^{i \Theta_j}$ operators. See appendix \ref{constrchiplusapp} for a calculation of this type.}

A few other comments about $\chi_+$: first, we would like to point out that the conditions (\ref{implicit_definition}) do not \emph{uniquely} determine $\chi_+$ in terms of $(K,W,\chi)$. In particular, if $\chi_+$ is a solution of (\ref{implicit_definition}) then $\chi_+' = \chi_+ + \delta$ is also a solution as long as $\delta$ satisfies the two conditions
\begin{align}
\Lambda_+^T\delta =0 \pmod{2}, \quad \quad W^T \delta = \delta, 
\label{delta_constraint}
\end{align}
where the first equality holds for all integer vectors $\Lambda_+$ with $W\Lambda_+=\Lambda_+$. That being said, these different choices of $\chi_+$ are not physically distinct and are akin to different gauge choices in electromagnetism. 

Another important comment is that there are general constraints on $\chi_+$ which are similar in spirit to the constraints (\ref{symmetry_constraints_1}) on $K, W$. More specifically, using the fact that $S^2 = 1$, one can show that $\chi_+$ must obey
\beq\label{chi_plus_constraint}
2\chi_+=\text{diag}(KW + K) \pmod 2
\eeq
(see appendix \ref{constrchiplusapp} for a proof).

Finally, we would like to mention an \emph{explicit} formula for $\chi_+$ in terms of $(K, W, \chi)$. To use this formula, one first has to make an integer change of basis so that $(K,W,\chi)$ are in a particular form. In particular, as we show in appendix \ref{standard_form}, it is always possible to put $(K,W, \chi)$ into the following standard form:
\begin{align}
W &=\bpm
-\mathbf{1_{n_- - m}}& 0 & 0 & 0\\
0 & \mathbf{1_{n_+-m}} & 0 & 0\\
0 & 0 & 0 & \mathbf{1_{m}}\\
0 & 0 &  \mathbf{1_{m}} & 0
\epm, \nonumber \\
K &=\bpm
A& 0 & B & -B\\
0 & C & D & D\\
B^T & D^T & E & F\\
-B^T & D^T & F^T & E
\epm, \quad 
\chi=\bpm
 0 \\
 \chi_2 \\
 0 \\
 0
 \epm \label{k_final_maintext}
\end{align}
Here, in the first equation, $\mathbf{{1}_m}$ denotes an $m \times m$ identity matrix, and $m, n_+, n_-$ are non-negative integers satisfying $n_+ + n_- = N$ and $m \leq n_\pm$. In the second equation, $A, C, E, F$ are symmetric integer square matrices with the same dimensions as the diagonal 
blocks of $W$. Finally, in the third equation, $\chi_2$ is a column vector with 
$ n_+ - m$ components, all of which are \emph{integer}.

Once $K, W$ and $\chi$ are in the above standard form, then $\chi_+$ is given by (see appendix \ref{chiplusformapp})
\beq\label{chi_plus_general}
\chi_+=
\bpm
0 \\
\chi_2 + 2a\\
\text{diag}(E+F)/2 + b\\
\text{diag}(E+F)/2 + b
\epm   
\eeq
Here $a,b$ are arbitrary integer vectors that parameterize the different choices of $\chi_+$, as discussed in Eq.~(\ref{delta_constraint}).

\subsection{The $K_\pm$ matrices}
In addition to the auxiliary vector $\chi_+$, we also define two auxiliary matrices, $K_+$ and $K_-$. Roughly speaking, these matrices describe the restriction of $K$ to the $+1$ and $-1$ eigenspaces of $W$. More precisely, $K_\pm$ are defined as follows. First, define two sets, $\Xi_+$ and $\Xi_-$, by
\begin{equation}
\Xi_\pm = \{ \Lambda_{\pm} :\ W\Lambda_{\pm} = \pm \Lambda_{\pm}, \ \Lambda_{\pm} \in \mathbb{Z}^{N}\}
\end{equation}
These two sets form integer lattices of dimension $n_+$ and $n_-$, where $n_\pm$ are the dimensions of the $\pm 1$ eigenspaces of $W$. It follows that $\Xi_\pm$ can be represented as $\Xi_\pm = V_\pm \mathbb{Z}^{n_{\pm}}$ where $V_\pm$ are $N \times n_\pm$ integer matrices. We then define the matrices $K_\pm$ as
\begin{equation}
K_\pm = V_\pm^T K V_\pm
\label{Kplusminusdef}
\end{equation}

For a more concrete definition of $K_+$ and $K_-$, suppose that $(K,W,\chi)$ are in the standard form (\ref{k_final_maintext}). Then, $V_\pm$ can be written as
\begin{align}
V_+ = \bpm 0 & 0 \\
	   \mathbf{1_{n_+ - m}} & 0 \\
	   0 & \mathbf{1_{m}} \\
	   0 & \mathbf{1_{m}} \epm, \quad 
V_- = \bpm \mathbf{1_{n_- - m}} & 0 \\
	   0 & 0 \\
	   0 & \mathbf{1_{m}} \\
	   0 & -\mathbf{1_{m}} \epm,
\label{vpmgen}
\end{align}
so $K_\pm$ are given by
\begin{align}
K_+ = \bpm C & 2D \\ 2D^T & 2(E+F) \epm, \quad  K_- = \bpm A & 2B \\ 2B^T & 2(E-F) \epm
\label{kpmform}
\end{align}

\subsection{SPT criterion}
With this background, we are now ready to state our criterion for when a chiral boson edge theory $(K,W,\chi)$ can be gapped without breaking the $\mz_2$ symmetry. We begin by describing the criterion for the special case of edge theories of symmetry-protected topological (SPT) phases. These SPT edge theories are characterized by two properties:
\begin{align}
|\det(K)| = 1, \quad \text{sig}(K) = 0
\end{align}
Here the first property comes from the fact that SPT phases do not support non-trivial anyon excitations, while the second property comes from the fact that SPT phases have a vanishing chiral central charge: $c_- = 0$.\footnote{Both of these properties follow from the definition of SPT phases given in Ref.~\onlinecite{ChenGuLiuWen}.}

Our criterion for SPT edge theories is as follows: an SPT edge theory $(K,W,\chi)$ can be gapped without breaking the $\mz_2$ symmetry if and only if
\begin{align}
\nu = 0 \pmod{2}
\label{SPT_crit}
\end{align}
where $\nu$ is defined by
\begin{align}
\nu \equiv \frac{1}{2} \chi_+^T K^{-1} \chi_+ +\frac{1}{4}\text{sig}(K(1-W)) \pmod{2}
\end{align}
Here, $\chi_+$ is the auxiliary vector defined in Eq.~(\ref{implicit_definition}), and `$\text{sig}$' stands for signature.\footnote{The matrix $K(1-W)$ also has vanishing eigenvalues, but we define the signature by considering only the eigenvalues that are strictly positive and strictly negative.} 

To understand this criterion, it is important to recognize a few properties of $\nu$ (see appendices \ref{nu_is_ind_of_chi} and \ref{nu_is_integer} for proofs):

\begin{itemize}
\item $\nu$ takes the same value for \emph{any} $\chi_+$ obeying (\ref{implicit_definition}).

\item $\nu$ is always an integer in the bosonic SPT case, and is a multiple of $1/4$ in the fermionic SPT case.
\end{itemize}

The significance of the first property of $\nu$ is that it ensures that $\nu$ is \emph{well-defined} --- that is, $\nu$ does not depend on an arbitrary choice of $\chi_+$. As for the second property, notice that it tells us that $\nu$ can take $2$ distinct values in the bosonic case and $8$ values in the fermionic case. At the same time, recall that there are believed to be $2$ distinct bosonic SPT phases\cite{ChenLiuWen, ChenGuLiuWen} and $8$ distinct fermionic SPT phases,\cite{RyuZhang, Qiz8, YaoRyu, gu2014effect} which have a $\mz_2$ and $\mz_8$ group structure, respectively. Putting these two facts together, it follows that $\nu$ can be directly interpreted in terms of the $\mz_2$ ($\mz_8$) index of bulk bosonic (fermionic) SPT phases.

\subsection{General criterion} \label{gencritsect}
Having warmed up with the SPT case, we can now state our general criterion for chiral boson edge theories with $\mz_2$ symmetry. In fact, we will present two versions of this criterion with different advantages and disadvantages.

The first version of the criterion is that an edge theory $(K,W,\chi)$ can be gapped without breaking the $\mz_2$ symmetry if and only if three conditions are satisfied:
\begin{enumerate}[(I)]
\item The chiral central charge vanishes: $\sig(K)=0$.
\item The set of anyons contains a Lagrangian subgroup $\mathcal{L}$ that is invariant under the $\mz_2$ symmetry.
\item Let $U$ be an $N \times N$ integer matrix with the property that $U \mathbb{Z}^N = \Gamma$ where
$\Gamma$ is the anyon lattice corresponding to the Lagrangian subgroup $\mathcal{L}$: $\Gamma = \{l+ K \Lambda : \ l \in \mathcal{L}, \ \Lambda \in \mathbb{Z}^N \}$. Then there exists at least one choice of $\chi_+$ satisfying Eq.~(\ref{implicit_definition}) such that $\tilde{\chi}_+ \equiv U^T K^{-1} \chi_+$ obeys
\begin{align}
\frac{1}{2} \tilde{\chi}_+^T \tilde{K}^{-1} \tilde{\chi}_+ +\frac{1}{4}\text{sig}(\tilde{K}(1-\tilde{W})) = 0 \pmod{2} \label{tildenu=0} \\
2 \tilde{\chi}_+ = \text{diag}(\tilde{K} + \tilde{K} \tilde{W}) \pmod{2} \label{tildechiconst}  
\end{align} 
where $\tilde{K} \equiv U^T K^{-1} U$ and $\tilde{W} \equiv U^{-1} W^T U$.
\end{enumerate}

Let us explain the physical interpretation of the above conditions. Conditions (I) and (II) are easy to understand. Condition (I) is equivalent to requiring that there are an equal number of left and right moving edge modes. This is obviously a necessary condition for having a gapped edge since an edge theory with different numbers of left and right moving modes can never be gapped. Condition (II) is also straightforward: regardless of symmetry concerns, it is known that any Abelian topological phase with a gapped edge must have at least one Lagrangian subgroup.\cite{levin2013protected} This Lagrangian subgroup can be physically interpreted as the set of anyons that can be annihilated/absorbed at the gapped boundary. Clearly this set must be invariant under the symmetry if the boundary is $\mz_2$ symmetric.

Condition (III) is more subtle. To understand this condition, one needs to make two observations. The first observation is that if we were to \emph{condense} the anyons in the Lagrangian subgroup $\mathcal{L}$ then this anyon condensation\cite{Burnellanyoncondensation} would drive a bulk phase transition from an SET phase to a phase with no anyons, i.e. an \emph{SPT} phase. The second observation is that the three {quantities $(\tilde{K}, \tilde{W}, \tilde{\chi}_+)$ can} be interpreted as the data describing the edge theory for this SPT phase. 

If we accept the above interpretation, we see that Eq.~(\ref{tildenu=0}) is simply the requirement that the condensed SPT phase is \emph{trivial} (\ref{SPT_crit}). Likewise, Eq.~(\ref{tildechiconst}) is equivalent to the general self-consistency condition (\ref{chi_plus_constraint}) on the SPT edge theory. Putting this all together, we conclude that condition (III) is equivalent to requiring that there exists at least one\footnote{For each Lagrangian subgroup, there are generally multiple ways to condense the associated anyons while maintaining the $\mz_2$ symmetry, since the condensed anyons can carry different $\mz_2$ quantum numbers.\cite{Burnelldoublesemion2016, QiJianWang} In our formalism, these different condensation patterns are parameterized by different choices of $\chi_+$ satisfying Eq.~(\ref{implicit_definition}) and (\ref{tildechiconst}).} way to condense the anyons in the Lagrangian subgroup $\mathcal{L}$ while maintaining the $\mz_2$ symmetry such that the result is a trivial SPT phase. 

 
We now discuss the second version of the criterion. In this version, an edge theory $(K,W,\chi)$ can be gapped without breaking the $\mz_2$ symmetry if and only if the following four conditions are satisfied:
\begin{enumerate}[(i)]
\item The chiral central charge vanishes: $\sig(K)=0$.
\item The set of anyons contains a Lagrangian subgroup $\mathcal{L}$ that is invariant under the $\mz_2$ symmetry.
\item $|\det({K_\pm})| \cdot 2^{|\text{sig}(K_\pm)|}$ is a perfect square. 
\item There exists at least one choice of $\chi_+$ obeying Eq.~(\ref{implicit_definition}) such that $g(\chi_+) = 0 \pmod{2}$, where
\begin{align}
\label{general_invariant}
g(x) \equiv \frac{1}{2} x^T K^{-1} x +\frac{1}{4}\text{sig}(K(1-W)) 
\end{align}
\end{enumerate}
We can see that the only difference from the first version of the criterion is that condition (III) has been replaced by (iii) and (iv). The advantage of this replacement is that conditions (iii) and (iv) are easier to check in practice; the disadvantage is that their physical meaning is less clear. 

Indeed, the physical interpretation of condition (iii) is mysterious to us. As for condition (iv), the only insight we can provide relies on a conjecture. This conjecture states that the following two sets are identical:
\begin{align}
\{e^{\pi i g(\chi_+)}: \chi_+ \text{ obeying } (\ref{implicit_definition})\} = \{e^{2 i \theta_a}: \text{$\mz_2$ fluxes $a$} \} 
\end{align}
Here the set on the left hand side is defined by letting $\chi_+$ run over all solutions of (\ref{implicit_definition}). The set on the right hand side is defined via the `gauging' construction: imagine coupling the SET phase with edge theory $(K,W,\chi)$ to a dynamical $\mz_2$ gauge field. Then let $a$ run over all the $\mz_2$ flux excitations in this \emph{gauged} SET phase and let $e^{i \theta_a}$ denote the topological spin of $a$. 

If the above conjecture is correct then condition (iv) is equivalent to requiring that the gauged SET phase has at least one $\mz_2$ flux excitation with topological spin $e^{i \theta_a} = \pm 1$. The latter condition can be motivated by noting that it is a weaker version\footnote{Any anyon $a$ that belongs to a Lagrangian algebra necessarily satisfies $e^{i \theta_a} = 1$.} of condition (3) from the criterion in the introduction. This condition has also been previously conjectured to be a necessary requirement for an SET phase to support a gapped symmetric edge.\cite{LuVishwanathz2sl} 

A few more comments about conditions (i)-(iv): first, we should point out that while condition (iv) looks very similar to the SPT criterion discussed above, there is an important difference between the two: in the case of general edge theories, $g(\chi_+)$ does not necessarily take the same value for every $\chi_+$ obeying Eq.~(\ref{implicit_definition}). Therefore, to check condition (iv) it is necessary to compute $g(\chi_+)$ for \emph{all} $\chi_+$ obeying Eq.~(\ref{implicit_definition}), not just a single choice, as in the SPT case. 

Another important comment is that conditions (i)-(iv) are not all independent of one another in the case of bosonic systems. In particular, one can show that condition (iii) is guaranteed to hold if conditions (i) and (ii) hold.\footnote{In appendix \ref{K_plusminus_null_vectors}, we show that if conditions (i) and (ii) hold and $\text{sig}(K_\pm) = 0$ then $K_\pm$ have a complete set of null vectors. A corollary of this result is that $|\det{K_\pm}|$ is a perfect square under these conditions. This corollary is enough to prove the claim, since we can always ensure that $\text{sig}(K_\pm) = 0$ by stacking with an appropriate number of copies of the edge theory (\ref{K22_3}) which has $|\det{K_\pm}| = 2$ and $\text{sig}(K_\pm) = \pm 1$.} This means that we can omit condition (iii) in the bosonic case. 


\subsection{Examples}

We now illustrate both the SPT criterion and the general criterion with a few examples. 

\subsubsection{SPT examples}
{\bf Example 1:}
The first example we consider is the bosonic SPT edge theory with
\beq\label{K22_1}
K =
\bpm
0 & 1 \\
1 & 0 
\epm,\ \ 
W =
\bpm
1 & 0 \\
0 & 1 
\epm,\ \
\chi =
\bpm
1  \\
0  
\epm
\eeq
The corresponding $\mz_2$ symmetry transformation is
\begin{align}
S^{-1} e^{i\Theta_1} S = -e^{i\Theta_1}, \quad \quad S^{-1} e^{i\Theta_2} S = e^{i\Theta_2}
\end{align}
It is easy to see that the above edge theory can be gapped without breaking the $\mz_2$ symmetry: indeed, the perturbation $U \cos(\Theta_2)$ does the job, according to the conditions described in section \ref{gappertsect}. Thus, this edge theory can be identified with the boundary of a \emph{trivial} SPT phase.\cite{levin2012braiding} 

Let us check that our criterion gives the same result. To do this, we first need to compute $\chi_+$. Applying the formula in Eq.~(\ref{chi_plus_general}), we see that one choice for $\chi_+$ is
\begin{equation}
\chi_+ = \chi = \bpm 1 \\ 0 \epm
\end{equation}
Computing $\nu$, we obtain
\begin{align}
\nu = \frac{1}{2} \cdot 0 + \frac{1}{4} \cdot 0 = 0 \pmod{2}
\end{align}
Thus, our criterion correctly predicts that the edge theory (\ref{K22_1}) can be gapped without breaking the symmetry.
\\
\\
{\bf Example 2:}
Next we consider the bosonic SPT edge theory with
\beq\label{K22_2}
K =
\bpm
0 & 1 \\
1 & 0 
\epm,\ \ 
W =
\bpm
1 & 0 \\
0 & 1 
\epm,\ \
\chi =
\bpm
1  \\
1 
\epm
\eeq
The corresponding $\mz_2$ symmetry transformation is:
\begin{align}
S^{-1} e^{i\Theta_1} S = -e^{i\Theta_1}, \quad \quad S^{-1} e^{i\Theta_2} S = -e^{i\Theta_2}
\end{align}
In contrast to the first example, it is known that this edge theory cannot be gapped without breaking the $\mz_2$ symmetry. Thus the edge theory (\ref{K22_2}) can be identified with the boundary of a \emph{non-trivial} SPT phase.\cite{levin2012braiding} 

Again, let us check that our criterion gives the right result. Using the formula in Eq.~(\ref{chi_plus_general}), we have
\begin{equation}
\chi_+ = \chi = \bpm 1 \\ 1 \epm
\end{equation}
Hence
\begin{equation}
\nu = \frac{1}{2} \cdot 2 + \frac{1}{4} \cdot 0 = 1 \pmod{2}
\end{equation}
Thus, our criterion correctly predicts that the edge theory (\ref{K22_2}) cannot be gapped without breaking the symmetry.
\\
\\
{\bf Example 3:}
The third example we consider is the following bosonic SPT edge theory:\footnote{The edge theory (\ref{K22_3}) was first discussed in Ref.~\onlinecite{lu2012theory}, but was conjectured to be unphysical in that work.} 
\beq\label{K22_3}
K=
\bpm
0 & 1 \\
1 & 0 
\epm,\ \
W=
\bpm
0 & 1 \\
1 & 0 
\epm,\ \
\chi=
\bpm
0  \\
0  
\epm
\eeq
This data corresponds to a symmetry transformation that \emph{exchanges} the two electron operators, $e^{i \Theta_1}, e^{i \Theta_2}$: 
\begin{align}
S^{-1} e^{i\Theta_1} S = e^{i\Theta_2}, \quad \quad S^{-1} e^{i\Theta_2} S = e^{i\Theta_1}
\label{K22_3_electron}
\end{align}

It is easy to see that the above edge theory (\ref{K22_3}) does not support $\mz_2$ symmetric gapping perturbations of the null-vector type discussed in section \ref{gappertsect}. Indeed the null-vector condition (\ref{null_criterion}) requires that we consider perturbations of the form $U \cos( \Lambda^T \Theta - \alpha)$ where either the first or second component of $\Lambda$ is zero, while the symmetry condition (\ref{symmetry_criterion}) requires that the two components of $\Lambda$ are equal. Clearly these two conditions cannot be simultaneously satisfied for any non-trivial $\Lambda$. Nevertheless, in appendix \ref{0101}, we show that (\ref{K22_3}) \emph{can} be gapped using a different kind of $\mz_2$ symmetric perturbation, which is not of the null-vector type. The latter result implies the above edge theory can be identified with the boundary of a \emph{trivial} SPT phase.

Now let us check whether our criterion gives the right answer. According to Eq.~(\ref{chi_plus_general}), 
\beq
\chi_+=
\bpm
1/2 \\
1/2 
\epm
\eeq
Now computing $\nu$ we find
\begin{align}
\nu &=\frac{1}{2} \cdot \frac{1}{2} - \frac{1}{4} = 0 \pmod{2}
\label{nu=0}
\end{align}
Again, our criterion gives the correct prediction: the edge theory (\ref{K22_3}) can be gapped without breaking the symmetry.
\\
\\
{\bf Example 4:}
As a final example, consider the following \emph{fermionic} SPT edge theory:
\beq\label{K22_4}
K=
\bpm
1 & 0 \\
0 & -1 
\epm,\ \
W=
\bpm
-1 & 0 \\
0 & 1 
\epm,\ \
\chi=
\bpm
0  \\
0  
\epm
\eeq
This data corresponds to the symmetry transformation 
\begin{align}
S^{-1} e^{i\Theta_1} S = e^{-i\Theta_1}, \quad \quad S^{-1} e^{i\Theta_2} S &= e^{i\Theta_2}
\label{K22_4_electron}
\end{align}
The above edge theory describes the boundary of a non-trivial fermionic SPT phase --- in particular the `root' phase with $\mz_8$ index equal to $1$. To see this, let us refermionize the edge theory, and define two fermion operators $c_1^\dagger = e^{i \Theta_1}$, $c_2^\dagger = e^{i \Theta_2}$. The symmetry transformation then becomes
\begin{align}
S^{-1} c_1^\dagger S = c_1, \quad \quad S^{-1} c_2^\dagger S = c_2^\dagger
\end{align} 
If we then break the complex fermions $c_1, c_2$ into four Majorana fermions: $c_1 = \frac{1}{2}(\gamma_1 + i \gamma_2)$, and $c_2 = \frac{1}{2}(\gamma_3 + i \gamma_4)$, we then have
\begin{align}
S^{-1} \gamma_1 S &= \gamma_1, \quad \quad S^{-1} \gamma_2 S = -\gamma_2 \nonumber \\
S^{-1} \gamma_3 S &= \gamma_3, \quad \quad S^{-1} \gamma_4 S = \gamma_4 
\end{align} 
In this language, the edge consists of $4$ Majorana fermion modes, of which $\gamma_1, \gamma_2$ are right moving and $\gamma_3, \gamma_4$ are left-moving. It is clear that we can gap out the $\gamma_1$ and $\gamma_3$ mode by adding the symmetry-allowed backscattering term $i \gamma_1 \gamma_3$. The resulting edge theory then has one right-moving mode $\gamma_2$ and one left moving mode $\gamma_4$, where $\gamma_2$ is odd under the symmetry and $\gamma_4$ is even under the symmetry. This is exactly the edge theory of the fermionic SPT phase with $\mz_8$ index equal to $1$.\cite{RyuZhang, Qiz8, YaoRyu, gu2014effect}

We now check to see that our criterion gives the right result for this edge theory. According to Eq.~(\ref{chi_plus_general}),
\beq
\chi_+=
\bpm
0 \\
0 
\epm
\eeq
Now computing $\nu$ we find
\begin{align}
\nu &=\frac{1}{2} \cdot 0 + \frac{1}{4} = \frac{1}{4} \pmod{2}
\end{align}
Hence our criterion correctly predicts that the edge cannot be gapped. It also correctly predicts that the bulk SPT phase has a $\mz_8$ index equal to $1$ (the $\mz_8$ index is given by $ 4 \nu$ in the fermionic case).

\subsubsection{Examples of general criterion}
{\bf Example 1}:
As a first example of the \emph{general} criterion, consider the following set of edge theories:\cite{LuVishwanathz2sl}
\beq\label{K22_5}
K=
\bpm
0 & n \\
n & 0 
\epm,\ \
W=
\bpm
0 & 1 \\
1 & 0 
\epm,\ \
\chi=
\bpm
0  \\
0  
\epm ,
\eeq
where $n \geq 2$. These edge theories can be realized as boundaries of a particular class of bosonic SET phases. To understand the nature of these SET phases, notice that the formalism of section \ref{anyon_review} implies that there are $n^2$ different anyons in the bulk which can be parametrized as $l = (i,j)$ where $0\leq i,j \leq n-1$. The mutual statistics between two anyons is given by
\beq
\theta_{(i,j),(k,l)}=\frac{2\pi (il + jk)}{n}
\eeq
while the exchange statistics of a single anyon is given by
\beq
\theta_{(i,j)}=\frac{2\pi i j}{n}
\eeq
This braiding statistics data agrees exactly with the corresponding data for $\mz_n$ gauge theory where the anyon $l = (1,0)$ is the $\mz_n$ `charge' while $l = (0,1)$ is the $\mz_n$ `flux.' Thus, we can think of the bulk SET phase as a $\mz_n$ gauge theory. In the gauge theory language, we can think of the $\mz_2$ symmetry $W$ as an `electric-magnetic' symmetry that exchanges the charge and flux excitation, or more generally exchanges the anyons $(i,j) \leftrightarrow (j,i)$.\cite{BarkeshliWenzn}

We now apply our criterion to determine which of the above edge theories can by gapped without breaking the symmetry. The first step is to find the most general choice for $\chi_+$. The easiest way to do this is to use the explicit formula (\ref{chi_plus_general}), which gives
\beq
\chi_+ =
\bpm 
n/2 + b \\
n/2 + b 
\epm
\label{chiplusex1}
\eeq
for any integer $b$.

Having found $\chi_+$, the next step is to check conditions (i)-(iv) of the second version of the criterion. Clearly condition (i), i.e. $\text{sig}(K) = 0$, is satisfied for all $n$. Next consider condition (iii). To check this condition, we first compute $K_+$ and $K_-$ using the explicit formulas (\ref{kpmform}):
\begin{align}
K_+ = 2n, \quad K_- = -2n
\end{align}
Therefore
\begin{align}
|\det(K_\pm)| \cdot 2^{|\text{sig}(K_\pm)|} = 4n
\end{align}
so condition (iii) is satisfied if and only if $n$ is a perfect square.

As for condition (ii), it is easy to see that if $n$ is a perfect square then there always exists a Lagrangian subgroup invariant under the symmetry.  Indeed, the follow subgroup does the job:
\begin{equation*}
\mac{L}=\{ \ (k\cdot\sqrt{n},l\cdot\sqrt{n}) : \  0 \leq k,l < \sqrt{n} \ \}
\end{equation*}

Moving on to condition (iv), we now show that if $n$ is perfect square then there always exists a choice of $\chi_+$ such that $g(\chi_+) = 0 \pmod{2}$. Indeed, from (\ref{chiplusex1}) we have
\begin{equation}
g(\chi_+) = \frac{1}{2} \frac{(n+2b)^2}{2n} - \frac{1}{4}
\end{equation}
One can then see that this expression vanishes if we choose 
\beq
b=\frac{-n + \sqrt{n}}{2}
\eeq
Putting this all together, our criterion tells us that the SET edge theory (\ref{K22_5}) can be gapped without breaking the symmetry if and only if $n$ is a perfect square. 
\\
\\
{\bf Example 2}:
Another interesting set of examples is given by the following edge theories:
\beq\label{K22_6}
K=
\bpm
n & 0 \\
0 & -n 
\epm,\ \
W=
\bpm
-1 & 0 \\
0 & 1 
\epm,\ \
\chi=
\bpm
0  \\
\chi_2  
\epm ,
\eeq
where $n \geq 2$ and $\chi_2$ is an integer. Note that these edge theories describe the boundaries of bosonic SET phases when $n$ is even and fermionic SET phases when $n$ is odd.

Let us apply our criterion to determine which of the above edge theories can be gapped. The first step is to find the most general choice for $\chi_+$. The easiest way to do this is to use the explicit formula (\ref{chi_plus_general}), which gives
\beq
\chi_+ =
\bpm 
0 \\
\chi_2 + 2a 
\epm
\label{chiplusex2}
\eeq
for any integer $a$.

Now let us check conditions (i)-(iv) of the second version of our criterion (second version). Clearly condition (i), i.e. $\text{sig}(K) = 0$, is satisfied for all $n$. As for condition (iii),
we have
\begin{align}
K_+ = -n, \quad K_- = n
\end{align}
Therefore 
\begin{align}
|\det(K_\pm)| \cdot 2^{|\text{sig}(K_\pm)|} = 2n
\end{align}
It follows that condition (iii) is satisfied if and only if $n/2$ is a perfect square.

Moving on to condition (ii), it is not hard to see that if $n/2$ is a perfect square, then there is always a Lagrangian subgroup that is invariant under the symmetry:
\begin{align*}
\mac{L}=\{ \ \left(k\cdot\sqrt{n/2},l\cdot\sqrt{n/2} \right) : \  &0 \leq k,l < \sqrt{2n}, \\
								      &k+l = 0 \pmod{2} \ \}
\end{align*}

As for condition (iv), we now show that assuming $n/2$ is perfect square, then there exists a choice of $\chi_+$ such that $g(\chi_+) = 0 \pmod{2}$ if and only if $\chi_2 = \sqrt{n/2} \pmod{2}$. To see this, note that Eq.~(\ref{chiplusex2}) gives
\begin{equation}
g(\chi_+) = -\frac{1}{2} \frac{(\chi_2+2a)^2}{n} + \frac{1}{4}
\end{equation}
One can then see that this expression vanishes if we choose 
\beq
a=\frac{\sqrt{n/2}-\chi_2}{2}
\eeq
Conversely, it is not hard to check that there is no solution to $g(\chi_+) = 0 \pmod{2}$ if $\chi_2 \neq \sqrt{n/2} \pmod{2}$.
Putting this all together, our criterion tells us that the SET edge theory (\ref{K22_6}) can be gapped without breaking the symmetry if and only if $n/2$ is a perfect square and $\chi_2 = \sqrt{n/2} \pmod{2}$. 
\\
\\
{\bf Example 3}:
As a final example, we now discuss a subset of chiral boson edge theories which are especially easy to analyze using our criterion: namely edge theories with $W=\mathbbm{1}$. In this case the $\mz_2$ symmetry does not permute any anyons, so our criterion simplifies in several ways. First, because the anyons are not permuted, any Lagrangian subgroup is automatically invariant under the symmetry. Also, it is easy to see that $K_+ = K$ and $K_- = \emptyset$ in this case, so condition (iii) reduces to the requirement that $|\det(K)|$ is perfect square. The latter condition is guaranteed to hold as long there is a Lagrangian subgroup, so we can drop condition (iii) entirely. Finally, using Eq.~(\ref{chi_plus_general}), we see that the most general choice for $\chi_+$ is $\chi_+ = \chi + 2a$ where $a$ is an integer vector. Putting this all together, we conclude that this class of SETs has a gapped symmetric edge if and only if (a) $\sig(K) = 0$; (b) the set of anyons has a Lagrangian subgroup; and (c)
\begin{equation}
\frac{1}{2} (\chi^T + 2a^T) K^{-1} (\chi + 2a) = 0 \pmod{2}
\end{equation}
for some integer vector $a$. 




\section{Derivation for bosonic SPT edge theories}\label{boseSPTder}
In this section we prove our criterion for the special case of bosonic SPT edge theories. We start with this special case both because the derivation is simpler and because we will need these results when we prove the general criterion in section \ref{SETder}.

\subsection{Preliminary simplifications}\label{simplifications}
Our goal is to establish the criterion (\ref{SPT_crit}): that is, a bosonic SPT edge theory $(K,W,\chi)$ can be gapped if and only if $\nu = 0 \pmod{2}$. We begin with a few simplifications. First, recall that $\nu$ can only take the values $0$ or $1$ $\pmod{2}$ in the bosonic case (see Appendix \ref{nu_is_integer}). Given this fact, it suffices to prove two claims: 
\begin{itemize}
\item
If $\nu = 0 \pmod{2}$, the edge theory $(K,W,\chi)$ can be gapped.

\item
If $\nu = 1 \pmod{2}$, the edge theory $(K,W,\chi)$ cannot be gapped.
\end{itemize}

To simplify the proof further, we observe that $\nu$ is additive under `stacking' of edge theories: that is if $\{K_a,W_a,\chi_a\}$ and $\{K_b,W_b,\chi_b\}$ are two edge theories, and $\{K_{a+b},W_{a+b},\chi_{a+b}\}$ denotes their direct sum, then the corresponding $\nu$'s obey 
\beq
\nu_{a+b} =\nu_a+\nu_b
\label{add_prop}
\eeq
Given this observation, we now argue that it suffices to prove the first claim: that is, the second claim follows for free. To see this, let $(K,W,\chi)$ be a bosonic SPT edge theory with $\nu = 1 \pmod{2}$ and consider the composite edge theory obtained by taking a direct sum of $(K,W,\chi)$ and the edge theory (\ref{K22_2}). By the additivity property (\ref{add_prop}), this composite edge theory has $\nu = 0 \pmod{2}$ since the edge (\ref{K22_2}) has $\nu = 1$. Hence, if we can prove the first claim, then it follows that this composite edge theory can be gapped, which means that the corresponding bulk SPT phase is trivial. This in turn implies that the SPT phase associated with $(K,W,\chi)$ is non-trivial (since stacking it with a non-trivial SPT phase gives a trivial SPT phase and bosonic SPT phases obey a $\mz_2$ group law under stacking). Hence the edge theory $(K,W,\chi)$ cannot be gapped, establishing the second claim.

In fact, it suffices to prove the first claim in the special case
\begin{align}
\text{sig}(K(1-W)) = 0
\end{align}
One way to see this is to recall the edge theory (\ref{K22_3}). This is the edge theory of a trivial bosonic SPT phase, as we prove in Appendix \ref{0101}. Furthermore, this theory has $\nu = 0$. Thus, stacking with this edge theory does not affect either the gappability of the edge or the value of $\nu$. At the same time, the theory (\ref{K22_3}) has $\text{sig}(K(1-W)) = -1$, so given any edge theory, we can always stack with an appropriate number of copies of the edge theory (\ref{K22_3}) so as to ensure that the combined theory has $\text{sig}(K(1-W)) = 0$.


\subsection{Derivation}\label{proof_section_1}
As discussed in section \ref{simplifications}, it suffices to show that we can gap the bosonic SPT edge theories $(K,W,\chi)$ with $\nu = 0 \pmod{2}$ and $\text{sig}(K(1-W)) = 0$. We will do this by showing that such edge theories always support a set of linearly independent integer vectors $\{\Lambda_{+}^{(1)},...,\Lambda_+^{(n_+/2)}, \Lambda_-^{(1)},...,\Lambda_-^{(n_-/2)}\}$ with four properties: 
\begin{enumerate}

\item
The vectors are mutually null: $(\Lambda_{\pm}^{(j)})^T K \Lambda_{\pm}^{(k)} = 0$.

\item
The vectors are eigenvectors of $W$: $W \Lambda_{\pm}^{(j)} = \pm \Lambda_{\pm}^{(j)}$.

\item
$\{\Lambda_+^{(1)},...,\Lambda_+^{(n_+/2)}\}$ are primitive.

\item
$(\Lambda_+^{(j)})^T \chi_+ = 0 \pmod{2}$. 
\end{enumerate}
Using the above null vectors, we will construct a perturbation (\ref{perturbations}) that gaps the edge theory without breaking the $\mz_2$ symmetry. Here $n_+$ and $n_-$ denote the dimensions of the $+1$ and $-1$ eigenspaces of $W$. 

 \subsubsection{Proving $K_{\pm}$ have null vectors}\label{proof_section_3}
In this section, we show that both the $K_+$ and $K_-$ matrices (\ref{Kplusminusdef}) have a complete set of null vectors. We will see why this result is useful in section \ref{proof_section_4}, when we use the null vectors of $K_\pm$ to construct corresponding null vectors for $K$.

To establish that $K_+$ and $K_-$ both have a complete set of null vectors, it suffices to prove two claims: (i) $\text{sig}(K_\pm) = 0$, and (ii) the Abelian topological phases corresponding to the $K$-matrices $K_\pm$ have Lagrangian subgroups. Indeed, Ref.~\onlinecite{levin2013protected} showed that these two properties guarantee the existence of the required null vectors.

To see that $\text{sig}(K_\pm) = 0$, observe that 
\begin{align}
\text{sig}(K_-) &= \text{sig}(K(1-W)), \nonumber \\
\text{sig}(K_+) &= \text{sig}(K) - \text{sig}(K_-)
\end{align}
The fact that $\text{sig}(K_\pm) = 0$ then follows from our assumption that $\text{sig}(K(1-W)) = 0$, together with the fact that $\text{sig}(K) = 0$ for any SPT edge theory.

Proving that the topological phases corresponding to $K_\pm$ have Lagrangian subgroups requires a bit more work. We accomplish this using the following lemma (see below for proof):
\begin{lemma}
If a bosonic Abelian topological phase has the property that the chiral central charge $c_- = 0 \pmod{8}$, and that $a \times a = 1$ for every anyon $a$, then the phase has a Lagrangian subgroup. 
\label{lemma1}
\end{lemma}

With this lemma in hand, our task reduces to showing that (a) $K_\pm$ describe bosonic topological phases, and (b) all the anyons in these phases obey $a \times a = 1$. The first property will follow if we can show that $K_\pm$ have only even elements on the diagonal, while the second property will follow if we can show that $2 K_\pm^{-1}$ are integer matrices. 

The fact that $K_\pm$ have only even elements on the diagonal follows from the definition $K_\pm \equiv V_\pm^T K V_\pm$ (\ref{Kplusminusdef}), together with the fact that $K$ has only even elements on the diagonal. As for the fact that $2 K_\pm^{-1}$ are integer matrices, we will now prove this claim for $K_+$ --- the proof for $K_-$ is similar. Let $x$ be an $n_+$ component vector such that $K_+ x$ is an integer vector. Then $\Lambda_+^T K V_+ x$ is an integer for every $\Lambda_+ \in \Xi_+$. At the same time, it is clear that $\Lambda_-^T K V_+ x = 0$ for every $\Lambda_- \in \Xi_-$ since $V_+ x$ is even under $W$, while $\Lambda_-$ is odd under $W$. Noting that every integer vector $y$ can be written as a linear combination $y = \frac{1}{2}(\Lambda_+ + \Lambda_-)$ with $\Lambda_\pm \in \Xi_\pm$, we deduce that $2 y^T K V_+ x$ is always an integer for every integer vector $y$. It then follows that $2 V_+ x$ is an integer vector since $\det{K} = \pm 1$. Hence $2x$ must be an integer vector (since $2 V_+ x$ is clearly an element of $\Xi_+$). Thus, we have shown that if $K_+ x$ is an integer vector then $2 x$ is an integer vector. The claim follows immediately.

We now give the proof of Lemma \ref{lemma1}:
\begin{proof}
First we prove a weaker result: we show that any nontrivial topological phase obeying the conditions of the lemma must contain at least one anyon $b \neq 1$ which is a \emph{boson}. To see this, recall the general relation between the chiral central charge and the topological spins of the anyons in a bosonic topological phase:\cite{kitaevlongpaper}
 \beq
\frac{1}{\mathcal{D}} \sum_{a\in\mathcal{A}}d_a^2e^{i \theta_a}  = e^{2\pi i c_-/8},\ \ \ \mathcal{D}=\sqrt{\sum_{a\in\mathcal{A}} d_a^2}
 \eeq
Here $\mathcal{A}$ denotes the set of anyons in the (bosonic) topological phase, while $e^{i \theta_a}$ is the topological spin and $d_a$ is the quantum dimension of anyon $a$, and $c_-$ denotes the chiral central charge. 

In the case at hand, we have $c_-=0 \pmod{8}$ and we also know that $d_a=1$ for all $a$ because the phase is Abelian. Thus, this identity can be rewritten as
   \beq
\sum_{a\in\mathcal{A}}e^{ i\theta_a}  = \sqrt{|\mathcal{A}|}
\label{chicentid}
 \eeq
where $|\mathcal{A}|$ denotes the number of anyons in the topological phase. At the same time, since $a \times a = 1$ for every $a$, we know that the topological spins can only take $4$ possible values: $e^{i\theta_a} \in \{\pm 1, \pm i\}$ (this follows from the composition rule for exchange statistics $e^{i \theta_{a \times a}} = e^{4 i \theta_a}$ together with the fact that the topological spin is the same as exchange statistics in the Abelian case). Putting these two facts together, we see that if the topological phase is nontrivial i.e. $|\mathcal{A}| > 1$, then there must be at least one anyon $b \neq 1$ with $e^{i\theta_b} = 1$, since this is the only way the left hand side of (\ref{chicentid}) can have a real part larger than $1$. This proves that there is always at least one anyon $b \neq 1$ which is a boson.

With the help of the above result, we now prove the lemma. The argument is by induction: we assume the lemma holds for all topological phases with $n-1$ or fewer anyons, and we show that the lemma holds for phases with $n$ anyons. To this end, consider any topological phase with $n$ anyons that obeys the above conditions. As we argued above, this topological phase must contain at least one anyon $b \neq 1$ which is a boson. We can therefore \emph{condense} this boson and thereby construct a new topological phase with $n/4$ anyons. We will denote the set of anyons in this condensed phase by $\mathcal{A}'$. Formally, $\mathcal{A}'$ is defined as follows. Let $\mathcal{Z}_b$ be the set of all anyons in $\mathcal{A}$ that have trivial mutual statistics with respect to $b$. Then the set of anyons in the condensed phase $\mathcal{A}'$ is given by the quotient group $\mathcal{A}' =\mathcal{Z}_b/\{1,b\}$. Following this formal definition, it is easy to see that the condensed phase also obeys $a \times a = 1$ for all $a \in \mathcal{A}'$. One can also show that the condensed phase obeys $c_- = 0 \pmod{8}$.
Hence, by our inductive assumption the condensed phase must have a Lagrangian subgroup $\mathcal{L}$. One can then check that the subgroup of anyons generated by $\<\mathcal{L}, b\>$ gives a Lagrangian subgroup for the original phase. This completes the inductive step.
 \end{proof}

  
\subsubsection{Proving $K$ has symmetric null vectors}\label{proof_section_4}
So far we have shown that $K_\pm \equiv V_\pm^T K V_\pm$ each have a complete set of null vectors. Let us denote the null vectors of $K_+$ by 
$\{\bar{\Lambda}^{(1)}_+,...,\bar{\Lambda}^{(n_\pm/2)}_+\}$ and the null vectors of $K_-$ by $\{\bar{\Lambda}^{(1)}_-,...,\bar{\Lambda}^{(n_-/2)}_-\}$. 

The next step is to use these null vectors to construct corresponding null vectors for $K$: to do this, we simply define
\begin{equation}
\Lambda_{\pm}^{(j)}=V_{\pm}\bar{\Lambda}^{(j)}_{\pm}
\label{converteq}
\end{equation}
By construction, $\{\Lambda_{+}^{(1)},...,\Lambda_+^{(n_+/2)}, \Lambda_-^{(1)},...,\Lambda_-^{(n_-/2)}\}$ are mutually null with respect to $K$. Furthermore, it is clear that $W \Lambda_{\pm}^{(j)} = \pm \Lambda_{\pm}^{(j)}$. Thus, these vectors automatically satisfy the first two properties listed in the beginning of section \ref{proof_section_1}. The goal of this section is to show that they obey the third and fourth properties as well --- that is $\{\Lambda_+^{(1)},...\Lambda_+^{(n_+/2)}\}$ are primitive and $(\Lambda_+^{(j)})^T \chi_+ = 0 \pmod{2}$. 

We will do this by showing that we can always choose the $\bar{\Lambda}^{(j)}_+$ vectors so that $\{\bar{\Lambda}_+^{(1)},...\bar{\Lambda}_+^{(n_+/2)}\}$ are primitive and
\beq\label{symmetric_condition}
(\bar{\Lambda}_+^{(j)})^T \bar{\chi}_+=0 \pmod 2
\eeq
where $\bar{\chi}_+ = V_+^T \chi_+$.

We establish these two results using the following lemma (see below for proof):
\begin{lemma}
Let $K$ be an $n \times n$ symmetric, integer matrix and let $\chi$ be an $n$ component vector. Suppose $K, \chi$ are of the form
\begin{equation}
K = \bpm \kappa &  0 & 0 \\ 0 & 0 & 1 \\ 0 & 1 & 0 \epm, \quad \chi = \bpm x \\ 1 \\ 0 \epm
\label{Kchiblock}
\end{equation}
for some $(n-2) \times (n-2)$ matrix $\kappa$ and some $(n-2)$ component vector $x$. If (i) $K$ has $n/2$ linearly independent null vectors $\{\Lambda_1,...,\Lambda_{n/2}\}$ and (ii) $\frac{1}{2}\chi^T K^{-1} \chi = 0 \pmod{2}$, then there
exist another set of $n/2$ linearly independent, primitive null vectors $\{\Lambda_1',...,\Lambda_{n/2}'\}$ with $(\Lambda_i')^T \chi = 0 \pmod{2}$. 
\label{lemma2}
\end{lemma}

To apply the lemma in our case, we set $K=K_+$, $\chi=\bar{\chi}_+$. We can readily verify that $K_+, \bar{\chi}_+$ obey all the conditions of the lemma. Indeed, $K_+$ is clearly a symmetric, integer matrix. Furthermore one can check that $\frac{1}{2}\bar{\chi}_+^T K^{-1}_+ \bar{\chi}_+ = 0 \pmod{2}$: this follows from the identity
\begin{align}
\bar{\chi}_+^T K^{-1}_+ \bar{\chi}_+ = \chi_+^T K^{-1} \chi_+ 
\end{align}
together with the relation
\begin{align}
\frac{1}{2}\chi_+^T K^{-1} \chi_+ = 0 \pmod{2}, 
\end{align}
where the latter comes from our assumption that $\nu = 0 \pmod{2}$ and $\text{sig}(K(1-W)) = 0$. As for the condition that $K_+, \bar{\chi}_+$ have the block diagonal structure shown in Eq.~(\ref{Kchiblock}), we can assume this structure without loss of generality, using a `stacking' argument. In particular, we can always consider the direct sum of whatever edge theory we are interested in, and the trivial SPT edge theory (\ref{K22_1}). This stacking with the trivial SPT edge theory does not affect either the value of $\nu$ or the gappability of the edge (since (\ref{K22_1}) has $\nu = 0$), but it ensures that the composite edge theory has a $K_+$ and $\bar{\chi}$ of the form in (\ref{Kchiblock}).

We now give the proof of Lemma \ref{lemma2}:
\begin{proof}

To begin, let $m = \frac{1}{4}x^T \kappa^{-1} x$. By assumption $m$ is an integer. Next, define $w$ to be the following
$n$ component integer vector:
\begin{equation}
w = \det(\kappa) \cdot \bpm \kappa^{-1} x \\ -2m \\ 1 \epm
\end{equation}
The vector $w$ has two important properties. First, $w$ obeys $w^T K w = 0$. Second, $w$ has the property that 
\begin{equation}
K w =  \det(\kappa) \cdot \left[\chi - \bpm 0 \\ \vdots \\ 0 \\ 2m \epm \right]
\end{equation}
The latter property means that if $\Lambda$ is an integer vector with $\Lambda^T K w = 0$, then we will automatically 
have 
\begin{equation}
\Lambda^T \chi = 0 \pmod{2}
\label{lambdaprop}
\end{equation}
This is very useful since ultimately we want to find null vectors satisfying Eq.~(\ref{lambdaprop}).

The next step is to use the fact that $K$ has $n/2$ linearly independent null vectors $\Lambda_1,...,\Lambda_{n/2}$. Let $\mathcal{V}$ be the real-linear subspace defined by
\begin{equation}
\mathcal{V} = \{v: \ v = \sum_i a_i \Lambda_i + a w, \ \ v^T K w = 0, \ \ a_i, a \in \mathbb{R}\}
\end{equation}
It is easy to see that $\mathcal{V}$ is at least $n/2$ dimensional. To see this, note that either $w$ is
linearly independent from $\Lambda_1,...,\Lambda_{n/2}$ or it is linearly dependent. In the former case, $\mathcal V$ has $n/2+1$ generators
and $1$ constraint, while in the latter case, it has $n/2$ generators and no constraint since in this case $v^T K w=0$ follows from
the null property of the $\Lambda_i$'s. Thus, in either case $\mathcal{V}$ has dimension $n/2$. 

Given that $\mathcal{V}$ is defined by integer constraints ($v^T Kw = 0$) and integer generators ($\Lambda_i, w$), and given that it is 
$n/2$ dimensional, it follows that $\mathcal{V} \cap \mathbb{Z}^{n}$ is an $n/2$ dimensional integer lattice. Let $\{\Lambda_1',...,\Lambda_{n/2}'\}$
be generators of this lattice. By construction these vectors are linearly independent and primitive, and 
have the null property $(\Lambda_i')^T K \Lambda_j' = 0$. Furthermore,
$(\Lambda_i')^T K w = 0$ so $\Lambda_i'$ obeys Eq.~(\ref{lambdaprop}).

\end{proof}

\subsubsection{Constructing a gapping perturbation}\label{SSB}
At this point, we have constructed $N/2$ linearly independent null vectors of $K$, denoted $\{\Lambda_+^{(1)},...,\Lambda_+^{(n_+/2)}, \Lambda_-^{(1)},...,\Lambda_-^{(n_-/2)}\}$ satisfying the four properties listed in section \ref{proof_section_1}:
\begin{enumerate}

\item
$(\Lambda_{\pm}^{(j)})^T K \Lambda_{\pm}^{(k)} = 0$ for all $j,k$.

\item
$W \Lambda_{\pm}^{(j)} = \pm \Lambda_{\pm}^{(j)}$.

\item
$\{\Lambda_+^{(1)},...,\Lambda_+^{(n_+/2)}\}$ are primitive.

\item
$(\Lambda_+^{(j)})^T \chi_+ = 0 \pmod{2}$ for all $j$. 
\end{enumerate}

In this section, we use these null vectors to construct a perturbation that gaps the edge but does not break the $\mathbb{Z}_2$ symmetry explicitly or spontaneously.

Naively one might think that we could simply use the gapping perturbation $H'$ (\ref{perturbations}) with $\{\Lambda_i\} = \{\Lambda_\pm^{(i)}\}$. Indeed, properties 2 and 4 guarantee that this perturbation is invariant under the $\mz_2$ symmetry (for an appropriate choice of $\alpha_i$) and furthermore property (1) guarantees that it gaps the edge for sufficiently large $U$. The problem is that the full set of vectors $\{\Lambda_\pm^{(i)}\}$ is not necessarily primitive, i.e. it does not necessarily obey Eq.~(\ref{primcond}). As a result, we cannot rule out the possibility that the perturbation breaks the symmetry \emph{spontaneously}. To address this loophole, we now show how to construct a closely related perturbation using primitive null vectors $\{\Lambda_1,...,\Lambda_{N/2}\}$.

To begin, let $\mathcal{U}$ be the real subspace spanned by $\{\Lambda_{\pm}^{(i)}\}$:
\begin{equation}
\mathcal{U} = \{u : \ u = \sum_i c_i \Lambda_+^{(i)} + \sum_i d_i \Lambda_-^{(i)}, \ \ c_i, d_i \in \mathbb{R}\}
\end{equation}
Let $\Upsilon = \mathcal{U} \cap \mathbb{Z}^N$. Clearly $\Upsilon$ is an $N/2$ dimensional integer lattice. Furthermore, $\Upsilon$ is invariant under $W$ since $W \Lambda_\pm^{(i)} = \pm \Lambda_\pm^{(i)}$. Therefore, if we pick a basis for $\Upsilon$, then the action of $W$ on the basis vectors defines an $(N/2) \times (N/2)$ integer matrix that squares to the identity. Now, in appendix \ref{standard_form} we show that any integer matrix that squares to the identity can be put in the simple canonical form (\ref{w_final}). Applying this argument to $W$, it follows that we can find another basis for $\Upsilon$, denoted $\{\Lambda_1,...,\Lambda_{N/2}\}$ such that the action of $W$ on the $\Lambda_i$'s takes the form
\begin{align}
W \Lambda_{2i-1} &= \Lambda_{2i}, \quad W \Lambda_{2i} = \Lambda_{2i-1}, \quad 1 \leq i \leq q \nonumber \\
W \Lambda_i &= - \Lambda_i, \quad 2q+1 \leq i \leq 2q+r  \nonumber \\
W \Lambda_i &= \Lambda_i, \quad 2q+r+1 \leq i \leq N/2 
\end{align}
for some non-negative integers $q,r$ with $2q+r \leq N/2$. 

With the help of the $\Lambda_i$ vectors, we can now write down the desired gapping perturbation:
\begin{align}
H' &= \sum_{i=1}^q \int_{-\infty}^{\infty} \left[U \cos(\Lambda_{2i-1}^T \Theta) + U \cos(\Lambda_{2i}^T \Theta -\alpha_{2i}) \right] dx \nonumber\\
 &+ \sum_{i=2q+1}^{2q+r} \int_{-\infty}^{\infty} U \cos(\Lambda_{i}^T \Theta - \alpha_i) dx \nonumber \\
 &+ \sum_{i=2q+r+1}^{N/2} \int_{-\infty}^{\infty} U \cos(\Lambda_{i}^T \Theta) dx
\label{specpert}
\end{align}
Here the $\alpha_{2i}$ phases in the first term are fixed by requiring that
\begin{align}
e^{i\Lambda_{2i-1}^T \Theta} \rightarrow e^{i\Lambda_{2i}^T \Theta} \cdot e^{-i \alpha_{2i}}, \quad 1 \leq i \leq q
\end{align}
under the $\mathbb{Z}_2$ symmetry. Likewise, the $\alpha_i$ phases in the second term are fixed by requiring that
\begin{align}
e^{i\Lambda_{i}^T \Theta} \rightarrow e^{-i\Lambda_{i}^T \Theta} \cdot e^{2i\alpha_{i}}, \quad 2q+1 \leq i \leq 2q+r
\end{align}

We now explain why the above perturbation (\ref{specpert}) has the required properties: (a) it gaps the edge for sufficiently large $U$; (b) it does not break the $\mz_2$ symmetry explicitly; and (c) it does not break the symmetry spontaneously. To see that it gaps the edge, note that the $\Lambda_i$'s obey $\Lambda_i^T K \Lambda_j = 0$ since they are linear combinations of the $\Lambda_\pm^{(i)}$'s. To see that the perturbation does not break the $\mz_2$ symmetry explicitly, note that the first two terms of (\ref{specpert}) are manifestly invariant under the $\mz_2$ symmetry, so we only have to worry about the last term, $\sum_{i=2q+r+1}^{N/2} U \cos(\Lambda_{i}^T \Theta)$. To analyze the symmetry transformation properties of this term, note that $W \Lambda_i = \Lambda_i$ in this case so this term transforms as
\begin{align}
e^{i\Lambda_i^T \Theta} \rightarrow e^{i\Lambda_i^T \Theta} \cdot  e^{i \pi \Lambda_i^T \chi_+}, \quad 2q+r+1 \leq i \leq N/2
\end{align}
Thus, the crucial question is to determine the parity of $\Lambda_i^T \chi_+$. To this end, note that each of the above $\Lambda_i$'s can be written as a linear combination, 
$\Lambda_i = \sum_j a_{ij} \Lambda_+^{(j)}$ since $W \Lambda_i = \Lambda_i$. Furthermore, the expansion coefficients, $a_{ij}$, are all integers since the set $\{\Lambda_+^{(i)}\}$ is primitive (property 3). 
Therefore, since $(\Lambda_+^{(i)})^T \chi_+$ is even (property 4), we conclude that $\Lambda_i^T \chi_+$ is also even. Hence, the term $\sum_{i=2q+r+1}^{N/2} U \cos(\Lambda_{i}^T \Theta)$ is invariant under the $\mz_2$ symmetry, as required.

All that remains is to show that the above perturbation (\ref{specpert}) does not break the $\mz_2$ symmetry spontaneously. This follows from the fact that the set $\{\Lambda_1,...,\Lambda_{N/2}\}$ is primitive, which in turn follows from the fact that these vectors span the lattice $\Upsilon = \mathcal{U} \cap \mathbb{Z}^N$.\footnote{If a set of linearly independent integer vectors spans a lattice of the form $\mathcal{X} \cap \mathbb{Z}^N$ where $\mathcal{X}$ is a real-linear subspace, then it follows that the set is primitive.}

\section{Derivation for fermionic SPT edge theories}\label{fermSPTder}
In this section, we prove the criterion (\ref{SPT_crit}) in the \emph{fermionic} case: that is, we show that a fermionic SPT edge theory $(K,W,\chi)$ can be gapped if and only if $\nu = 0 \pmod{2}$. As in the bosonic case, we first simplify our task with a few observations. First, we note that $\nu$ can only take the values $\{0,1/4,...,7/4\}$ modulo $2$ (see appendix \ref{nu_is_integer}), and therefore it suffices to prove two claims: 
\begin{itemize}
\item
If $\nu = 0 \pmod{2}$, the edge theory $(K,W,\chi)$ can be gapped.

\item
If $\nu = 1/4,...,7/4 \pmod{2}$, the edge theory $(K,W,\chi)$ cannot be gapped.
\end{itemize}
Next we note that it suffices to prove the first claim: that is, the second claim follows for free. Indeed, this follows by the same stacking argument as in the bosonic case, but with the edge theory (\ref{K22_4}) taking the place of $(\ref{K22_3})$. 

The last simplification is to note that we can assume without loss of generality that
\begin{align}
\text{sig}(K(1-W)) = 0
\end{align}
The justification for this assumption is the same as in the bosonic case: we can guarantee that $\text{sig}(K(1-W)) = 0$ if we stack an appropriate number of copies of the edge theory (\ref{K22_3}) on top of the edge theory of interest.

Having made these simplifications, our task reduces to showing that any fermionic SPT edge theory with $\nu = 0 \pmod{2}$ and $\text{sig}(K(1-W)) = 0$ can be gapped without breaking the $\mz_2$ symmetry. This claim can be proved using \emph{exactly} the same arguments as in the bosonic case, with only one change: a new argument is needed to establish that $K_+$ and $K_-$ have a complete set of null vectors since the argument given in section \ref{proof_section_3} is not applicable. 

We now fill in this hole and present the argument for why $K_+$ and $K_-$ have a complete set of null vectors in the fermionic case. The first step is to note that by a result of Ref.~\onlinecite{levin2013protected}, it suffices to show that $K_\pm$ have two properties: (i) $\text{sig}(K_\pm) = 0$ and (ii) the topological phases corresponding to $K_\pm$ have Lagrangian subgroups. Property (i) is easy to establish: it follows immediately from the fact that $\text{sig}(K) = 0$ and $\text{sig}(K(1-W)) = 0$. Property (ii) is harder. To establish this, we use the following lemma (see below for proof):

\begin{lemma}
If a fermionic Abelian topological phase has the property that the number of distinct anyons is a perfect square and that $a \times a = 1$ for every anyon $a$, then the topological phase has a Lagrangian subgroup. 
\label{lemma3}
\end{lemma}

To apply this lemma in our case, we need to establish three points: (a) $K_\pm$ has at least one odd element on the diagonal, (b) $|\det(K_\pm)|$ is a perfect square and (c) $2 K_\pm^{-1}$ is an integer matrix. To establish (a), we note that we can always ensure that $K_+$ has at least one odd element on the diagonal by stacking with the following trivial SPT edge theory: $K = \bpm 1 & 0 \\ 0 & -1 \epm$, $W = \bpm 1 & 0 \\ 0 & 1 \epm$,
$\chi = \bpm 0 \\ 0 \epm$. This stacking does not affect either the gappability of the edge or the value of $\nu$, so it is harmless for our purposes. Likewise, we we can ensure that $K_-$ has at least one odd element on the diagonal by stacking with $K = \bpm 1 & 0 \\ 0 & -1 \epm$, $W = \bpm -1 & 0 \\ 0 & -1 \epm$, $\chi = \bpm 0 \\ 0 \epm$. To establish (b), we use a mathematical result, proved in appendix \ref{perfectsqapp}, which states that $|\det(K_\pm)| \cdot 2^{|\text{sig}(K_\pm)|}$ is always a perfect square for any SPT edge theory $(K,W,\chi)$ that has $\nu = 0 \pmod{1/2}$. Applying this result to our case, we immediately deduce that $|\det(K_\pm)|$ is a perfect square since $\text{sig}(K_\pm) = 0$. As for property (c), this follows from the same argument as in the bosonic case (section \ref{proof_section_3}).

To complete the argument, we now present the proof of Lemma \ref{lemma3}:

\begin{proof}
As in the bosonic case, we first prove a weaker result: we show that any non-trivial fermionic topological phase with the above properties supports at least one non-trivial anyon which is either a boson or a fermion. To see this, notice that all the anyons in such a topological phase must have mutual statistics $e^{i \theta_{a,b}} = \pm 1$ since they obey $a \times a = 1$. It then follows that all the anyons have exchange statistics $e^{i\theta_a} \in \{\pm 1, \pm i\}$. At the same time, it is easy to see that if two distinct anyons have exchange statistics $\pm i$, then fusing these anyons together gives a nontrivial anyon which is either a boson or a fermion: this follows from the composition rule $\theta_{a \times b} = \theta_a + \theta_b + \theta_{a,b}$. Combining these two observations, we see that the only way to avoid having a non-trivial anyon which is a boson or a fermion is if there is only \emph{one} non-trivial anyon. But this cannot be the case, since the number of distinct anyons is a perfect square, by assumption.

With the help of the above result, we now prove the lemma. Like in the bosonic case, the argument is by induction: we assume the lemma holds for all topological phases with $n-1$ or fewer anyons, and we show that the Lemma holds for phases with $n$ anyons. To this end, consider any topological phase with $n$ anyons which obeys the above conditions. As we argued above, this topological phase must contain at least one anyon $b$ which is a boson or a fermion. In fact, we can assume without loss of generality that $b$ is a boson, since we can always combine it with the local fermion if this is not the case. We can now imagine \emph{condensing} the boson $b$ and thereby constructing a new topological phase $\mathcal{A}'$ with $n/4$ anyons. Formally, the condensed phase $\mathcal{A}'$ is given by the quotient group $\mathcal{A}' =\mathcal{Z}_b/\{1,b\}$ where $\mathcal{Z}_b$ is the set of all anyons in $\mathcal{A}$ that have trivial mutual statistics with respect to $b$. Following this formal definition, it is easy to see that the condensed phase also obeys $a \times a = 1$ for all $a \in \mathcal{A}'$. Hence, by our inductive assumption the condensed phase must have a Lagrangian subgroup $\mathcal{L}$. As in the bosonic case, one can check that the subgroup of anyons generated by $\<\mathcal{L}, b\>$ gives a Lagrangian subgroup for the original phase. This completes the inductive step.

\end{proof}


\section{Derivation of general criterion}\label{SETder}

Having established our criterion for the special case of SPT edge theories, we now derive the two versions of the general criterion presented in section \ref{gencritsect}. We will do this in three steps: (1) we prove that the second version of the criterion is \emph{sufficient} for having a gapped edge; (2) we show that the first version of the criterion is \emph{necessary} for having a gapped edge; (3) we show that the first version of the criterion implies the second version.
 
\subsection{Sufficiency of second criterion} \label{suff_crit_sect}
We begin by proving the sufficiency of the second version of our criterion: in other words, we show that conditions (i)-(iv) are sufficient to guarantee that an edge theory $(K,W,\chi)$ can be gapped without breaking the $\mz_2$ symmetry. Our proof is brief since it uses almost identical arguments to the SPT case.

The first step is to observe that we can assume without loss of generality that $\text{sig}(K(1-W)) = 0$. The reason we can make this assumption is the same as in the SPT case: we can always guarantee that $\text{sig}(K(1-W)) = 0$ by stacking some number of copies of the edge theory (\ref{K22_3}) on top of the edge theory of interest. This stacking is harmless because it does not affect the gappability of the edge or conditions (1)-(4) of the criterion (since (\ref{K22_3}) is the edge theory for a trivial SPT phase).

The next step is to show that the matrices $K_+$ and $K_-$ have a complete set of null vectors. Unfortunately, the argument from the SPT case (e.g. sections \ref{proof_section_3} and \ref{fermSPTder}) does not generalize easily since it relies on the assumption that $|\det(K)| = 1$. Therefore, a new argument is necessary. We give this  argument in Appendix \ref{K_plusminus_null_vectors}; the proof uses conditions (i)-(iii) of the criterion together with our assumption that $\text{sig}(K(1-W)) = 0$. 

Let us denote the null vectors of $K_\pm$ by $\{\bar{\Lambda}_{\pm}^{(1)},...,\bar{\Lambda}_{\pm}^{(n_{\pm}/2)}\}$. The next step is to convert these null vectors for $K_\pm$ into null vectors for $K$. We do this by defining $\Lambda_\pm^{(j)} = V_\pm \bar{\Lambda}_\pm^{(j)}$ as in Eq.~(\ref{converteq}). By construction, $\{\Lambda_{\pm}^{(1)},...,\Lambda_{\pm}^{(n_{\pm}/2)}\}$ are mutually null with respect to $K$ and obey $W \Lambda_\pm^{(j)} = \pm \Lambda_\pm^{(j)}$. 

To proceed further, we note that condition (iv) of our criterion guarantees that we can always choose the above null vectors so that $\{\Lambda_+^{(1)},...,\Lambda_+^{(n_+/2})\}$ are primitive and $(\Lambda_+^{(j)})^T \chi_+ = 0 \pmod{2}$: indeed, this can be shown using the same argument as in the SPT case (section \ref{proof_section_4}). 

Putting this all together, we have established the existence of a complete set of null vectors $\{\Lambda_{\pm}^{(1)},...,\Lambda_{\pm}^{(n_{\pm}/2)}\}$ that obey $W \Lambda_\pm^{(j)} = \pm \Lambda_\pm^{(j)}$ and that have the property that $\{\Lambda_+^{(1)},...,\Lambda_+^{(n_+/2)}\}$ are primitive and $(\Lambda_+^{(j)})^T \chi_+ = 0 \pmod{2}$. The last step is to use these null vectors to devise an explicit perturbation $\sum_{i=1}^{N/2} U \cos(M_i^T \Theta -\alpha_i)$ that gaps the edge. Conveniently, this perturbation can be constructed in exactly the same way as the SPT case (section \ref{SSB}), without any changes. This completes the proof.

\subsection{Necessity of first criterion} \label{nec_crit_sect}

In this section we prove the necessity of the first version of our criterion. More specifically, we focus on proving the necessity of condition (III), since the necessity of conditions (I) and (II) was discussed in section \ref{gencritsect}. 

\begin{figure}[tb]
\includegraphics[width=.99\columnwidth]{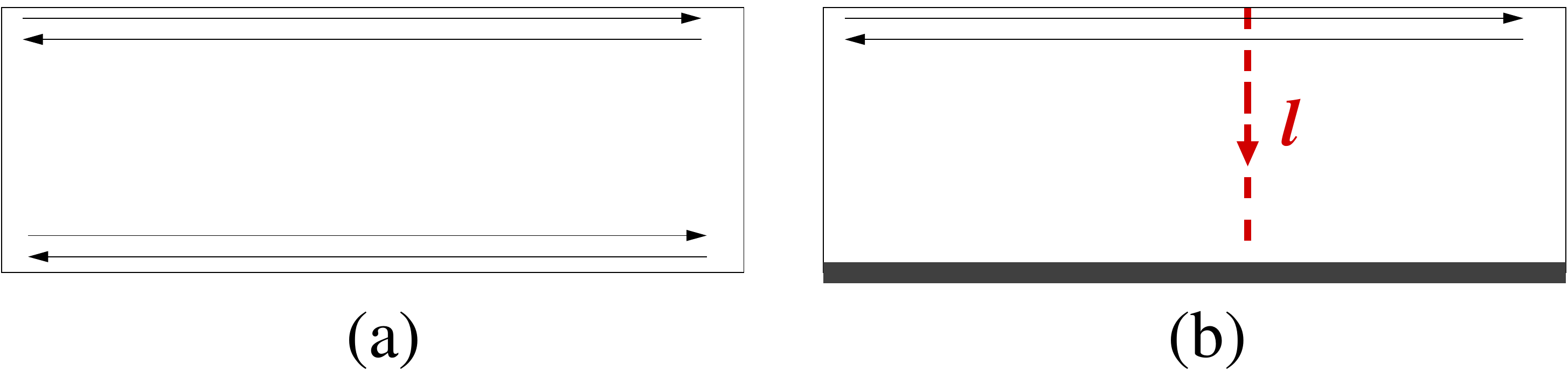}
\centering
\caption{Starting point for our argument: (a) We consider a 2D SET phase in a strip geometry in which both the upper and lower edge are described by the edge theory $(K,W,\chi)$. (b) We then imagine gapping the lower edge with appropriate interactions. The low energy theory for the strip is then similar to the edge theory for the upper edge $(K,W,\chi)$ except that it contains an additional set of local operators that describe processes where an anyon $l$ tunnels from the upper edge to the lower edge and is then annihilated (dashed line).}
\label{fig:tun_oper}
\end{figure}

\subsubsection{Outline of argument}

Let $(K,W,\chi)$ be an edge theory that can be gapped without breaking the $\mz_2$ symmetry. We need to show that condition (III) holds. To this end, we consider the 2D SET phase  corresponding to $(K,W,\chi)$ in a quasi one-dimensional strip geometry. We assume both the upper edge and the lower edge of the strip are described by the edge theory $(K,W,\chi)$ but with opposite chiralities (Fig. \ref{fig:tun_oper}a). Next, we imagine adding appropriate interactions to the lower edge so as to gap it without breaking the $\mz_2$ symmetry (Fig. \ref{fig:tun_oper}b); this is possible by our assumption. 

With this setup, our proof proceeds in three steps. In the first step, we show that the low energy theory for this strip is an $N$-component chiral boson theory with appropriately modified parameters $(\tilde{K}, \tilde{W}, \tilde{\chi})$. Next, we use the fact that the strip is a \emph{one-dimensional} system, and can therefore be thought of as the boundary of a trivial SPT phase, to deduce that the associated $\mz_2$ index vanishes: $\tilde{\nu} = 0 \pmod{2}$. We then use the fact that $\tilde{\nu} = 0 \pmod{2}$ to show that condition (III) of our criterion is satisfied.

\subsubsection{Proof}

Following the above outline, we begin by deriving the low energy theory for the quasi one-dimensional strip shown in Fig. \ref{fig:tun_oper}b. Naively, one might think that this low energy theory is simply the chiral boson theory $(K,W,\chi)$ that describes the upper edge since the lower edge is gapped. However, this is not quite correct: the reason is that the two low energy theories have different sets of local operators. In particular, in the $(K,W, \chi)$ edge theory, the fundamental local operators (i.e. the operators from which all other local operators can be built) are of the form $e^{i \Lambda^T \Theta} \equiv e^{i \Lambda^T K \Phi}$ where $\Lambda$ is an integer vector. In contrast, in the strip geometry, the set of local operators is larger. To see this, recall that for any gapped edge of an Abelian topological phase, there is always a nonempty set of anyons that can be annihilated (absorbed) at the edge. In particular, this is true for the lower edge of the strip. The existence of these anyons means that the strip has an additional set of local operators\footnote{These tunneling operators are local in the 1D sense because they are supported within a finite interval along the direction parallel to the strip.} which describe processes where an anyon $l$ tunnels from the upper edge to the lower edge of the strip, and is then annihilated at the lower edge (Fig. \ref{fig:tun_oper}b). These tunneling operators are of the form $e^{i l^T \Phi}$, so all together, the set of local operators can be parameterized as $e^{i (\Lambda^T K + l^T) \Phi}$ where $\Lambda$ is an integer vector and $l$ runs over the set of integer vectors $\mathcal{L}$ containing the anyons that can be annihilated at the lower edge.\cite{levin2013protected}

Given that the set of local operators is larger than usual, the low energy theory for the strip does not obey the standard normalization conventions for a chiral boson edge theory. To fix this discrepancy, we renormalize the fields at the upper edge: we define
\begin{align}
\tilde{\Phi}_i = \sum_{j=1}^N (U^{-1} K)_{ij} \Phi_j
\end{align}
where $U$ is an $N \times N$ integer matrix with the property that $U \mathbb{Z}^N = \Gamma$ where $\Gamma$ is the lattice 
\begin{align*}
\Gamma = \{l + K \Lambda: \ l \in \mathcal{L}, \ \Lambda \in \mathbb{Z}^N \}. 
\end{align*}
With this change of variables, the local operators in the low energy theory can be parameterized as 
\begin{align}
e^{i \tilde{\Lambda}^T \tilde{\Theta}}
\label{locop2}
\end{align}
where $\tilde{\Lambda}$ is an integer vector, $\tilde{\Theta}_j \equiv \sum_k \tilde{K}_{jk} \tilde{\Phi}_k$, and
\begin{equation}
\tilde{K} \equiv U^T K^{-1} U
\end{equation}

Notice that our renormalization has achieved its goal: the local operators (\ref{locop2}) are now parameterized in the same way as in a usual chiral boson edge theory, with $\tilde{K}$ playing the role of the $K$-matrix. Furthermore, one can check that $\tilde{K}$ is an integer matrix (this follows from the fact that $\mathcal{L}$ is a Lagrangian subgroup), and the $\tilde{\Phi}$ fields obey commutation relations of the usual form:
\begin{equation}
[\tilde{\Phi}_i(x'), \partial_x \tilde{\Phi}_j(x)] = 2\pi i \tilde{K}_{ij}^{-1} \delta(x-x')
\end{equation}

So far we have worked out the operator content of the low energy theory; next, we analyze how operators transform under the $\mz_2$ symmetry. First, let us consider the subset of operators of the form $e^{i \tilde{\Lambda}^T \tilde{\Theta}}$ where $\tilde{\Lambda}$ is of the form $\tilde{\Lambda} = U^{-1} K \Lambda$ for some integer vector $\Lambda$. These operators are special because they are built solely out of electron creation/annihilation operators on the upper edge, and do not involve any anyon tunneling operators. Indeed, from the definitions, one can see that $e^{i \tilde{\Lambda}^T \tilde{\Theta}} = e^{i \Lambda^T \Theta}$. As a result, the $\mz_2$ transformation law for these operators is directly determined by Eq.~(\ref{z2symmetry_2}). In particular, we have
\begin{align}
S^{-1} e^{i \tilde{\Lambda}^T \tilde{\Theta}} S &= S^{-1} e^{i \Lambda^T \Theta} S \nonumber \\
& = e^{i \Lambda^T W^T \Theta} \cdot \text{(phase)} \nonumber \\
& = e^{i \tilde{\Lambda}^T \tilde{W}^T \tilde{\Theta}} \cdot \text{(phase)}
\end{align}
where
\begin{align}
\tilde{W} \equiv U^{-1} W^T U
\end{align}
and $(phase)$ denotes a phase factor which depends on $\chi$ and $\Lambda$, but whose explicit form is not important. From the above transformation law, we infer that the basic operators $e^{i \tilde{\Theta}_j}$ transform as
\begin{align}
S^{-1} e^{i \tilde{\Theta}_j} S &= e^{\sum_k \tilde{W}_{kj} \tilde{\Theta}_k} \cdot e^{i \tilde{\chi}_j}
\end{align}
for some phase factors $e^{i \tilde{\chi}_j}$. 

At this point, we have shown that the low energy theory for the strip looks like a standard chiral boson edge theory with parameters $(\tilde{K}, \tilde{W}, \tilde{\chi})$. The next step is to observe that the strip is a quasi \emph{one-dimensional} system. This means that we can think of the strip as the boundary of the vacuum --- i.e. the boundary of a trivial SPT phase. It then follows from the SPT criterion (\ref{SPT_crit}) that the $\mz_2$ index $\tilde{\nu}$ associated with this low energy theory must vanish: that is,
\begin{equation}
 \frac{1}{2}\tilde{\chi}_+^T \tilde{K}^{-1} \tilde{\chi}_+ + \frac{1}{4} \text{sig}(\tilde{K} (1-\tilde{W})) = 0 \pmod{2}
\label{tildenu}
\end{equation}
where $\tilde{\chi}_+$ is defined by the usual conditions
\begin{align}
S^{-1} e^{i \tilde{\Lambda}_+^T \tilde{\Theta}} S & = e^{ i \tilde{\Lambda}_+^T \tilde{\Theta}} \cdot e^{i\pi \tilde{\Lambda}_+^T \tilde{\chi}_+}, \nonumber \\
\tilde{W}^T \tilde{\chi}_+ &= \tilde{\chi}_+
\label{transtil} 
\end{align}
Here the first equality holds for all integer vectors $\tilde{\Lambda}_+$ that satisfy $ \tilde{W} \tilde{\Lambda}_+ =\tilde{\Lambda}_+$.

At the same time, the same consistency arguments that give the constraint (\ref{chi_plus_constraint}) imply that $\tilde{\chi}_+$ obeys
\begin{align}
2 \tilde{\chi}_+ = \text{diag}(\tilde{K} + \tilde{K} \tilde{W}) \pmod{2}
\label{tildechidiag}
\end{align}

With Eq.~(\ref{tildenu}) and (\ref{tildechidiag}) in hand, we are now ready to show that condition (III) of our criterion is satisfied. To this end, define
\begin{align}
\chi_+ \equiv K (U^{-1})^T \tilde{\chi}_+ 
\end{align}
We will argue that $\chi_+$ obeys all the requirements of condition (III). Indeed, by construction, $\tilde{\chi}_+ = U^T K^{-1} \chi_+$, and $\tilde{\chi}_+$ obeys Eqs.~(\ref{tildenu=0}-\ref{tildechiconst}). Thus, all that remains is to show that $\chi_+$ satisfies (\ref{implicit_definition}).
The first equation of (\ref{implicit_definition}), namely $S^{-1} e^{i \Lambda_+^T \Theta} S  = e^{ i \Lambda_+^T \Theta} \cdot e^{i\pi \Lambda_+^T \chi_+}$, can be verified straightforwardly. First, we note that $e^{i \Lambda_+^T \Theta} = e^{i \tilde{\Lambda}_+^T \tilde{\Theta}}$ where
$\tilde{\Lambda}_+ = U^{-1} K \Lambda_+$. It then follows that
\begin{align}
S^{-1} e^{i \Lambda^T_+ \Theta} S &= S^{-1} e^{i \tilde{\Lambda}^T_+ \tilde{\Theta}} S \nonumber \\
& = e^{i \tilde{\Lambda}^T_+ \tilde{\Theta}} \cdot e^{i \pi \tilde{\Lambda}^T_+ \tilde{\chi}_+} \nonumber \\
& = e^{i \Lambda^T_+ \Theta} \cdot e^{i \pi \Lambda^T_+ \chi_+}
\label{trans}
\end{align}
as we wanted to show (here the second equality follows from (\ref{transtil})). The second equation of (\ref{implicit_definition}), namely $W^T \chi_+ = \chi_+$, also follows from a straightforward calculation:
\begin{align}
W^T \chi_+ &= W^T K (U^{-1})^T \tilde{\chi}_+ \nonumber \\
&= K (U^{-1})^T \tilde{W}^T \tilde{\chi}_+ \nonumber \\
&= K (U^{-1})^T \tilde{\chi}_+ \nonumber \\
&= \chi_+
\end{align}

\subsection{First criterion implies second criterion}
To complete our derivation, we now show that the first version of the criterion implies the second version, or equivalently condition (III) implies conditions (iii)-(iv).

To show that (III) implies (iv), we use the following identities:
\begin{align}
\chi_+^T K^{-1} \chi_+  &= \tilde{\chi}_+^T \tilde{K}^{-1} \tilde{\chi}_+, \nonumber \\
\text{sig}( K (1-W)) &=  \text{sig}(\tilde{K} (1-\tilde{W}))
\end{align}
Combining these two equations gives 
\begin{align}
g(\chi_+) =  \frac{1}{2}\tilde{\chi}_+^T \tilde{K}^{-1} \tilde{\chi}_+ + \frac{1}{4} \text{sig}(\tilde{K} (1-\tilde{W}))
\end{align}
From this identity we can see immediately that (III) implies (iv).

All that remains is to show that (III) implies (iii). To this end, we use a lemma proved in appendix \ref{perfectsqapp} which states that if $\tilde{\chi}_+$ obeys Eq.~(\ref{tildenu=0}) and (\ref{tildechiconst}) then the following quantity is a perfect square:
\begin{align}
|\det(\tilde{K}_\pm)| \cdot 2^{|\text{sig}(\tilde{K}_\pm)|}
\label{pfsq}
\end{align}
Here $\tilde{K}_\pm$ is defined by
\begin{equation}
\tilde{K}_\pm \equiv \tilde{V}_\pm^T \tilde{K} \tilde{V}_\pm
\end{equation}
while $\tilde{V}_\pm$ is an $n_\pm \times n_\pm$ matrix with the property that $\tilde{V}_\pm \mz^{n_\pm} = \tilde{\Xi}_\pm$, where
\begin{equation}
\tilde{\Xi}_\pm = \{ \Lambda_{\pm} :\ \tilde{W} \Lambda_{\pm} = \pm \Lambda_{\pm}, \ \Lambda_{\pm} \in \mathbb{Z}^{N}\}
\end{equation}

Given that the quantity in (\ref{pfsq}) is a perfect square, it suffices to prove that
\begin{align}
\frac{|\det(K_\pm)| \cdot 2^{|\text{sig}(K_\pm)|}}{|\det(\tilde{K}_\pm)| \cdot 2^{|\text{sig}(\tilde{K}_\pm)|}} = m_\pm^2 
\label{finid}
\end{align}
where $m_\pm$ is an integer. Condition (iii) will then follow immediately.

To derive (\ref{finid}), the key observation is that $K \Xi_\pm \subset U \tilde{\Xi}_\pm$. It follows that there must exist integer matrices $R_\pm$, of dimension $n_\pm \times n_\pm$ such that $K V_\pm = U \tilde{V}_\pm R_\pm$. Hence
\begin{align}
K_\pm = R_\pm^T \tilde{K}_\pm R_\pm
\end{align}
so that 
\begin{align}
|\det(K_\pm)| = |\det(\tilde{K}_\pm)| \cdot |\det(R_\pm)|^2
\end{align}
At the same time, it is easy to see that
\begin{align}
\text{sig}(\tilde{K}_\pm) = \text{sig}(K_\pm) 
\end{align}
Putting this all together, we derive Eq.~(\ref{finid}), with $m_\pm = \det(R_\pm)$.


\section{Conclusion}

In this paper we have derived a necessary and sufficient criterion for when an SET phase with Abelian anyons and a unitary $\mz_2$ symmetry has a gapped symmetric edge. Our criterion is phrased in terms of chiral boson edge theories, and as such, it applies to any SET phase that is consistent with such an edge theory.

Perhaps the best way to understand our criterion is the interpretation given in section \ref{gencritsect}. According to this interpretation,
first version of our criterion states that an SET phase with $\mz_2$ symmetry can have a gapped symmetric edge if and only if there exists a way to condense a collection of anyons, while maintaining the $\mz_2$ symmetry, such that the result is a \emph{trivial} SPT phase. 

The advantage of this way of looking at our criterion is that it suggests natural generalizations to other symmetry groups as well as to non-Abelian phases. The main obstruction to implementing such generalizations is finding a convenient framework for describing the relevant class of SET phases that allows one to (a) systematically search for anyon condensation patterns that preserve the symmetry and (b) determine whether the SPT phase that is produced in each case is trivial or non-trivial. In this paper, we were able to achieve these goals using the chiral boson edge theory formalism but other frameworks are also possible. For example, SET phases with unitary onsite symmetry group $G$ can be described, quite generally, using mathematical structures known as `braided $G$-crossed categories.'\cite{BarkeshliBondersonChengWang,TeoHughesFradkin,TarantinoLindnerFidkowski} An interesting direction for future work would be to derive a criterion for protected edge modes in this language.


\begin{acknowledgments}
We thank Sav Sethi for bringing Ref.~\onlinecite{lecheminant2002} to our attention. CH and ML are supported in part by the NSF under grant No. DMR-1254741.
\end{acknowledgments}

\begin{appendix}

\section{General constraint on $\chi_+$} \label{constrchiplusapp}

In this appendix, we derive the general constraint on $\chi_+$ (\ref{chi_plus_constraint}) which we reprint below for convenience:
\begin{align}
2 \chi_+ = \text{diag}(K + KW) \pmod{2}
\end{align}
Our basic strategy is to work out how operators of the form 
\begin{align}
e^{i \Lambda^T \Theta} \cdot e^{i \Lambda^T W^T \Theta}
\end{align}
transform under the $\mz_2$ symmetry, and then compare with the definition of $\chi_+$ (\ref{implicit_definition}) . 

To this end, notice that Eq.~(\ref{z2symmetry_2}) implies that
\begin{align}
S^{-1} e^{i \Lambda^T \Theta(x)} S = e^{i \Lambda^T W^T \Theta(x)} \cdot e^{i \beta}
\end{align}
where $\beta$ is a real number that depends on $\Lambda$. At the same time, since $S^2 = 1$, we have
\begin{align}
S^{-1} e^{i \Lambda^T W^T \Theta(x)} S = e^{i \Lambda^T \Theta(x)} \cdot e^{-i \beta}
\end{align}
Multiplying these equations together gives 
\begin{align}
S^{-1} \left(e^{i \Lambda^T \Theta(x)} e^{i \Lambda^T W^T \Theta(x)} \right)S = e^{i \Lambda^T W^T \Theta(x)} e^{i \Lambda^T \Theta(x)} 
\label{eq1ap1}
\end{align}
since the two factors of $e^{\pm i \beta}$ cancel one another.

Next we use the Baker-Campbell-Hausdorff formula to reverse the order of the two terms on the right hand side of Eq.~(\ref{eq1ap1}):
\begin{align}
e^{i \Lambda^T W^T \Theta(x)} e^{i \Lambda^T \Theta(x)} = \ &e^{i \Lambda^T \Theta(x)} e^{i \Lambda^T W^T \Theta(x)} \nonumber \\
 & \cdot e^{[\Lambda^T \Theta(x), \ \Lambda^T W^T\Theta(x)]}
\label{eq2ap1}
\end{align}
To evaluate the commutator, we use the commutation relation
\begin{align}
[\Theta_i(x), \Theta_j(x')] = \pi i \left(K_{ij} \ \text{sgn}(x'-x) + X_{ij} \right)
\label{commklein}  
\end{align}
Here $X_{ij}$ is an anti-symmetric real matrix which is chosen so that $e^{i \Theta_i(x)}$ and $e^{i \Theta_j(x')}$ have the correct fermionic or bosonic commutation relations when $x \neq x'$.\footnote{$X_{ij}$ plays the same role as the well-known \emph{Klein factors}.} We will discuss the explicit form of $X_{ij}$ below; for now, we will proceed with $X_{ij}$ undetermined. Substituting Eq.~(\ref{commklein}) into (\ref{eq2ap1}), and using $\text{sgn}(0) = 0$, gives 
\begin{align}
e^{i \Lambda^T W^T \Theta(x)} e^{i \Lambda^T \Theta(x)} = \ & e^{i \Lambda^T \Theta(x)} e^{i \Lambda^T W^T \Theta(x)} \nonumber \\
& \cdot e^{\pi i \Lambda^T X W \Lambda}
\end{align}
Combining this identity with Eq.~(\ref{eq1ap1}) gives
\begin{align}
S^{-1} \left(e^{i \Lambda^T \Theta(x)} e^{i \Lambda^T W^T \Theta(x)} \right) S = \ & e^{i \Lambda^T \Theta(x)} e^{i \Lambda^T W^T \Theta(x)} \nonumber \\
&\cdot e^{\pi i \Lambda^T X W \Lambda}
\label{eq3ap1}
\end{align}

The next step is to compare the symmetry transformation (\ref{eq3ap1}) with the definition of $\chi_+$ (\ref{implicit_definition}). Consistency between these two equations implies the constraint
\begin{align}
(\Lambda^T + \Lambda^T W^T) \chi_+ = \Lambda^T X W \Lambda \pmod{2}
\end{align}
Equivalently, we can write this as
\begin{align}
2 \Lambda^T \chi_+ = \Lambda^T X W \Lambda \pmod{2}
\label{constraint1ap}
\end{align}
since $W^T \chi_+ = \chi_+$. 

At this point, we need to work out the explicit form for $X_{ij}$. To do this, we note that $X_{ij}$ should be chosen so that, for $x \neq x'$,
$e^{i \Theta_i(x)}$ and $e^{i \Theta_j(x')}$ anti-commute if $e^{i\Theta_i}$ and $e^{i\Theta_j}$ are both fermionic operators, and commute otherwise. Formally, this is equivalent to the condition that $e^{i \Theta_i(x)}$ and $e^{i \Theta_j(x')}$ anticommute if $K_{ii}$ and $K_{jj}$ are both odd and commute otherwise. Using the Baker-Campbell-Hausdorff formula, this reduces to the requirement
\begin{align}\label{defxij}
K_{ij} + X_{ij} = \begin{cases} 1 \pmod{2} & \text{if } K_{ii} \text{ and } K_{jj} \text{ both odd} \\
				0 \pmod{2} & \text{otherwise} \end{cases}
\end{align}
Equivalently, we have:
\begin{align}
X_{ij} = K_{ii} K_{jj} + K_{ij} \pmod{2}
\label{Xform}
\end{align}

Substituting Eq.~(\ref{Xform}) into (\ref{constraint1ap}) gives 
\begin{align}
2 \sum_i \Lambda_i \chi_{+i} = \sum_{ijk} \Lambda_i (K_{ii} K_{jj} + K_{ij}) W_{jk} \Lambda_k \pmod{2}
\end{align}
Next, we use the relation $\sum_j K_{jj} W_{jk} = K_{kk} \pmod{2}$ (which can be derived from $W^T K W = K$) to rewrite this constraint as
\begin{align}
2 \sum_i \Lambda_i \chi_{+i} = \sum_i \Lambda_i K_{ii} + \sum_{ik} \Lambda_i (KW)_{ik} \Lambda_k \pmod{2}
\label{constr2ap}
\end{align}
Finally, we note that since $KW$ is a symmetric matrix, $\sum_{ik} \Lambda_i (KW)_{ik} \Lambda_k = \sum_i \Lambda_i (KW)_{ii}$, modulo $2$. Substituting this into Eq.~(\ref{constr2ap}), we derive
\begin{align}
2 \chi_{+i} = K_{ii} + (KW)_{ii} \pmod{2}
\end{align}
as we wanted to show.


\section{Standard form for $K,W$ and $\chi$}\label{standard_form}

In this appendix, we show that given any chiral boson edge theory $(K,W,\chi)$, it is always possible to make an integer change of basis of the form
\begin{align}
W &\rightarrow U W U^{-1}, \nonumber \\
K &\rightarrow (U^{-1})^T K U^{-1}, 
\end{align}
where $U$ is an integer matrix with $\det(U) = \pm 1$, so that $K, W, \chi$ take the standard form shown in Eq.~(\ref{k_final_maintext}). Our derivation closely follows a similar argument in appendix A of Ref.~\onlinecite{levin2012classification}.

We start with $W$. To begin, we note that the fact that $W^2=1$ implies that the eigenvalues of $W$ are $\pm 1$. Let $n_+$ be the number $+1$ eigenvalues and $n_-$ be the number of $-1$ eigenvalues, so that $n_++n_-=N$. Next let $\{v_1,...,v_{n_+}\}$ be a basis for the $+1$ eigenspace. Notice that we can always choose the $v_i$ to be integer vectors, since the $+1$ eigenspace is spanned by the columns of $1+W$, an integer matrix. Furthermore, it is easy to see that we can always choose the $v_i$ so that they are primitive. This ensures that we can extend $\{v_1,\ldots,v_{n_+}\}$ to an integer basis $\{v_1,\ldots,v_{n_+},w_1,\ldots,w_{n_-}\}$ for the whole $N$ dimensional space such that the matrix with columns $\{v_1,\ldots,v_{n_+},w_1,\ldots,w_{n_-}\}$ has determinant $\pm 1$. 

Let $U^{-1}$ be the matrix with columns $\{v_1,\ldots,v_{n_+},w_1,\ldots,w_{n_-}\}$. We next make a change of basis $W\rightarrow UWU^{-1}$. After this change of basis $W$ is in the form
\beq
W=
\bpm
\mathbf{1_{n_+}} & F \\
0 & G
\epm
\eeq
where $F$ is $n_+\times n_-$ and $G$ is $n_-\times n_-$. Next, using the fact that $W^2=1$, we deduce that $G^2=1$. Furthermore, since $\text{Tr}(W)=n_+-n_-$, we know that $\text{Tr}(G)=-n_-$. We conclude that $G=-\mathbf{1_{n_-}}$ and thus $W$ is of the form
\beq
W=
\bpm
\mathbf{1_{n_+}} & F \\
0 & -\mathbf{1_{n_-}}
\epm
\eeq

The next step is to make another transformation $W\rightarrow UWU^{-1}$, where $U$ is an integer matrix of the form
\beq
U=
\bpm
U_1 & 0 \\
0 & U_2
\epm
\eeq
and $\det(U_1)=\det(U_2)=\pm 1$. Under this transformation $F\rightarrow U_1FU_2^{-1}$. Therefore, according to the Smith normal form for integer matrices, we can always choose $U_1$ and $U_2$ so that $F$ becomes a diagonal matrix. 

To proceed further, we make a transformation of the form $W\rightarrow UWU^{-1}$ where $U$ is of the form
\beq
U=
\bpm
\mathbf{1_{n_+}} & Y \\
0 & \mathbf{1_{n_-}}
\epm
\eeq
Under this transformation, $F\rightarrow F-2Y$. Hence we can choose $Y$ so that $F$ has only 0's and 1's along the diagonal. We can therefore assume without loss of generality that $F$ is of the form
\beq
F=
\bpm
\mathbf{1_{m}} & 0 \\
0 & 0
\epm
\eeq
where $m\le n_\pm$. At this point, we have managed to put $W$ in the form
\beq
W=\bpm
\mathbf{1_{m}}& 0 & \mathbf{1_{m}} & 0\\
0 & \mathbf{1_{n_+ - m}} & 0 & 0\\
0 & 0 & -\mathbf{1_{m}} & 0\\
0 & 0 &  0 & -\mathbf{1_{n_- - m}}
\epm
\eeq

The final step is to make another transformation $W\rightarrow UWU^{-1}$ where $U$ is of the form
\beq
U=\bpm
\mathbf{1_{m}}& 0 & 0 & 0\\
0 & \mathbf{1_{n_+ - m}} & 0 & 0\\
\mathbf{1_{m}} & 0 & \mathbf{1_{m}} & 0\\
0 & 0 &  0 & \mathbf{1_{n_- - m}}
\epm
\eeq
This transformation puts $W$ in the form
\beq
W=\bpm
0& 0 & \mathbf{1_{m}} & 0\\
0 & \mathbf{1_{n_+ - m}} & 0 & 0\\
\mathbf{1_{m}} & 0 &0& 0\\
0 & 0 &  0 & -\mathbf{1_{n_- - m}}
\epm
\eeq
Finally, after reordering the columns and rows we can put $W$ in the following form:
\beq\label{w_final}
W=\bpm
-\mathbf{1_{n_- -m}}& 0 & 0 & 0\\
0 & \mathbf{1_{n_+ - m}} & 0 & 0\\
0 & 0 & 0 & \mathbf{1_{m}}\\
0 & 0 &  \mathbf{1_{m}} & 0
\epm
\eeq
This is precisely the form for $W$ given in Eq.~(\ref{k_final_maintext}).

We now move on to consider $K$. Using the fact that $W^TKW=K$, and that $K$ is symmetric, we immediately deduce that $K$ must be of the form
\beq
K=\bpm
A& 0 & B & -B\\
0 & C & D & D\\
B^T & D^T & E & F\\
-B^T & D^T &  F^T & E
\epm
\eeq
where $A, C, E, F$ are symmetric integer matrices.

All that remains is to show that we can put $\chi$ in standard form. To this end, we note that since $S^2 = 1$, the vector $\chi$ must be of the form 
\begin{align}
\chi = \bpm \chi_1 \\ \chi_2 \\ \chi_3 \\ \chi_4 \epm
\end{align}
with $2\chi_2 = 0 \pmod{2}$ and $\chi_3 + \chi_4 = 0 \pmod{2}$. 

We now redefine the $\Theta$ fields by shifting them by a constant, that is $\Theta \rightarrow \Theta + \pi \alpha$ where
\begin{align}
\alpha = \bpm \alpha_1 \\ \alpha_2 \\ \alpha_3 \\ \alpha_4 \epm
\end{align}
Under this transformation
\begin{align}
\chi \rightarrow \chi + \bpm 2 \alpha_1 \\ 0 \\ \alpha_3 -\alpha_4 \\ \alpha_4 -\alpha_3 \epm
\end{align}
Clearly by choosing $\alpha$ appropriately, we can arrange so that $\chi_1 = 0$ and $\chi_3 = \chi_4 = 0 \pmod{2}$. Thus, after this redefinition, $\chi$ is of the form
\beq
 \chi=\bpm
 0 \\
 \chi_2 \\
 0 \\
 0
 \epm
 \eeq
where $\chi_2$ is an $(n_+ -m)$ component vector with integer components. This is exactly the standard form given in Eq.~(\ref{k_final_maintext}).


\section{Formula for $\chi_+$ in terms of $(K,W,\chi)$} \label{chiplusformapp}
In this appendix, we derive an explicit formula for $\chi_+$ (\ref{chi_plus_general}) which can be used whenever $(K,W,\chi)$ are in the standard form
(\ref{k_final_maintext}).

To begin, recall that the second condition in the definition of $\chi_+$ (\ref{implicit_definition}) is that $W^T \chi_+ = \chi_+$. Assuming $W$ is in the standard form (\ref{k_final_maintext}), this condition implies that
\beq
\chi_+=
\bpm
0 \\
x \\
y \\
y
\epm
\eeq
for some unknown vectors $x$ and $y$ of dimension $(n_+ - m)$ and $m$ respectively. Our task is to compute $x$ and $y$.

To compute $x$, let $\Lambda$ be an integer vector of the form 
\begin{align}
\Lambda = \bpm 0 \\ \Lambda_2 \\ 0 \\ 0 \epm
\end{align}
Substituting this $\Lambda$ into the definition of $\chi_+$ (\ref{implicit_definition}), we deduce that
\begin{align}
S^{-1} e^{i \Lambda^T \Theta} S = e^{i \Lambda^T \Theta} e^{i \pi \Lambda_2^T x}
\end{align}
At the same time, according to the general symmetry transformation (\ref{z2symmetry_2}), 
\begin{align}
S^{-1} e^{i \Lambda^T \Theta} S = e^{i \Lambda^T \Theta} e^{i \pi \Lambda_2^T \chi_2}
\end{align}
We conclude that
\begin{align}
x = \chi_2 + 2a
\end{align}
for some integer vector $a$. 

All that remains is to compute $y$. The most straightforward way to do this is to use the same approach as we did to calculate $x$, but with $\Lambda = \bpm 0 \\ 0 \\ \Lambda_3 \\ \Lambda_3 \epm$. However, we can find $y$ more quickly by leveraging the general constraint (\ref{chi_plus_constraint}) on $\chi_+$. Assuming $K, W$ are in the standard form (\ref{k_final_maintext}), this constraint immediately implies that
\begin{align}
2 y = \text{diag}(E+F) \pmod{2}
\end{align}
so that
\begin{align}
y = \text{diag}(E+F)/2 + b
\end{align}
for some integer vector $b$. (Note that, $y$ is only determined modulo $1$, unlike $x$, since the most general $\delta$ obeying conditions (\ref{delta_constraint}) is of the form $\delta = \bpm 0 & 2a & b & b \epm^T$.

Putting this all together, we have shown that
\begin{align}
\chi_+ = \bpm 0 \\ \chi_2 + 2a \\ \text{diag}(E+F)/2 + b \\ \text{diag}(E+F)/2 + b \epm
\end{align}
for some integer $a$ and $b$. This completes our derivation of the formula (\ref{chi_plus_general}).


\section{Proving $\nu$ is invariant under $\chi_+ \rightarrow \chi_+ + \delta$}\label{nu_is_ind_of_chi}

In this appendix, we show that $\nu$ does not depend on the choice of $\chi_+$ for any SPT edge theory. Equivalently, we show that $g(\chi_+ + \delta) = g(\chi_+) \pmod{2}$ where
\begin{align}
g(x) \equiv \frac{1}{2} x^T K^{-1} x +\frac{1}{4}\text{sig}(K(1-W)) 
\end{align}
and $\delta$ is any vector satisfying Eq.~(\ref{delta_constraint}).

The first step is to note that 
\begin{align}
g(\chi_+ + \delta) - g(\chi_+) &=  \frac{1}{2} \delta^T K^{-1} \delta + \chi_+^T K^{-1} \delta 
\end{align}
Next let $\Delta = K^{-1} \delta$. Then, we can rewrite the above difference as
\begin{align}
g(\chi_+ + \delta) - g(\chi_+) = \frac{1}{2} \Delta^T K \Delta + \chi_+^T \Delta
\label{nudiff}
\end{align}

To proceed further, we note that any $\delta$ satisfying (\ref{delta_constraint}) is necessarily of the form
\begin{equation}
\delta = (1+W^T) y
\label{delexp}
\end{equation}
where $y$ is an integer vector. Indeed, this is obvious if one works in the standard basis where $W$ is of the form (\ref{k_final_maintext}).

Next, we note that $\Delta$ can be parameterized in a similar manner to $\delta$, that is
\begin{equation}
\Delta = (1+ W) x
\label{delexp2}
\end{equation}
where $x$ is an integer vector. Indeed, this follows from (\ref{delexp}), using the identity $K^{-1} (1+W^T) = (1+W) K^{-1}$ together with the fact that $K^{-1}$ is an integer matrix in the SPT case.

Finally, we substitute (\ref{delexp2}) into (\ref{nudiff}) to obtain
\begin{align}
g(\chi_+ + \delta) - g(\chi_+) &= x^T K x + x^T K W x + 2 \chi_+^T x
\end{align}
It is easy to see that the sum of these three terms is always even, using the general constraint $2 \chi_+ = \text{diag}(KW + K) \pmod{2}$ (\ref{chi_plus_constraint}), together with the general identity $x^T M x + \text{diag}(M)^T x  = 0 \pmod{2}$ which holds for any symmetric integer matrix $M$ and integer vector $x$. Putting this all together, we conclude that 
\begin{align}
g(\chi_+ + \delta) = g(\chi_+) \pmod{2}, 
\end{align}
as claimed.


\section{Quantization of $\nu$}\label{nu_is_integer}

In this appendix we prove that $\nu$ is always an integer in the bosonic SPT case and always a multiple of $1/4$ in the fermionic SPT case. 
We begin with the bosonic case. Our proof is based on the following alternative expression for $8 \nu$: 
\begin{align}
8 \nu &= (2\chi_+)^T (K W)^{-1} (2 \chi_+) \nonumber \\
&+ 2 \cdot \text{sig}( K (1-W))
\label{8nu}
\end{align}
(Here we have used the fact that $W^T \chi_+ = \chi_+$). Given this expression, it suffices to show that the right hand side of (\ref{8nu}) is a multiple of $8$. We will accomplish this by finding a basis where $KW$ and $\chi_+$ are especially simple. 

To this end, consider the matrix $KW$. This matrix has several nice properties. First, it is \emph{symmetric} since $(KW)^T = W^T K = KW$. Second, it has determinant $\pm 1$, since $\det{K} = \pm 1$ in the SPT case. Finally, we can assume without loss of generality that $KW$ is \emph{odd} (i.e. has at least one odd element on the diagonal) and \emph{indefinite} (i.e. has at least one positive and one negative eigenvalue). Indeed, we can always ensure the latter two properties by stacking the edge theories (\ref{K22_3}) and (\ref{K22_1}) on top of the edge theory we are interested in, as this stacking only shifts $g(\chi_+)$ by an even integer.

To proceed further, we apply a theorem of Milnor, which states that if two integer matrices $M, M'$ satisfy the above properties and have the same dimension and same signature, then $M, M'$ are equivalent up to an integer change of basis: that is $M' = U M U^T$ for some integer matrix $U$ with determinant $\pm 1$.\cite{milnor1973symmetric} Applying Milnor's theorem to the matrix $K W$, it follows that there exists an integer basis where $KW$ is of the form
\begin{align}
KW = \bpm \pm 1 & 0 & 0 \cdots \\ 0 & \pm 1 & 0 & \cdots \\ 0 & 0 & \pm 1 & \cdots \\ \vdots & \vdots & \vdots & \cdots \epm
\end{align}

Next, we note that Eq.~(\ref{chi_plus_constraint}) implies that, in this basis, the vector $2 \chi_+$ has odd integer components. Combining this observation with the above expression for $KW$, we deduce that
\begin{align}
(2\chi_+)^T (KW)^{-1} (2 \chi_+) = \text{sig}(K W) \pmod{8}
\label{kwinv}
\end{align}
since $x^2 = 1\pmod{8}$ for any odd integer $x$. 

Comparing Eq.~(\ref{kwinv}) with the right hand side of (\ref{8nu}), it suffices to show that 
\begin{align}
\text{sig}(K W) + 2 \cdot \text{sig}(K (1-W))
\end{align}
is a multiple of $8$. The latter result follows from the two identities
\begin{align}
\text{sig}(K W) &= \text{sig}(K (1+W)) \nonumber \\
&- \text{sig}(K (1-W))  
\end{align}
and
\begin{align}
\text{sig}(K) &= \text{sig}(K (1+W)) \nonumber \\
&+ \text{sig}(K (1-W))  
\end{align}
together with the fact that $\text{sig}(K) = 0$ in the SPT case.

We now move on to the fermionic case. We need to show that $\nu$ is a multiple of $1/4$, or equivalently, $\frac{1}{2} \chi_+^T K^{-1} \chi_+$, is a multiple of $1/4$. To do this, we use the identity
\begin{align}
\frac{1}{2} \chi_+^T K^{-1} \chi_+ = \frac{1}{2} \bar{\chi}_+^T K_+^{-1} \bar{\chi}_+
\end{align}
where $\bar{\chi}_+ \equiv V_+^T \chi_+$. Next, we observe that $\bar{\chi}_+$ is an integer vector: this is obvious if one works in the standard basis where $\chi_+$ is of the form (\ref{chi_plus_general}) and $V_+$ is given by (\ref{vpmgen}). Also, $K_+^{-1}$ is a \emph{half}-integer matrix (see the argument given in section \ref{proof_section_3}). Putting this all together, it immediately follows that $\frac{1}{2} \chi_+^T K^{-1} \chi_+$ is a multiple of $1/4$ as we wanted to show.


\section{Proving the edge theory (\ref{K22_3}) can be gapped}\label{0101}

In this appendix, we show that the chiral boson edge theory (\ref{K22_3}) can be gapped without breaking the $\mz_2$ symmetry. We note that a similar argument was given in Ref.~\onlinecite{lecheminant2002}. 

The first step is to translate the $(K,W,\chi)$ data in Eq.~(\ref{K22_3}) into an explicit Hamiltonian,
\begin{align}
&H = \int_{-\infty}^{\infty} \frac{v}{4\pi} [(\partial_x \Theta_1)^2 + \frac{v}{4\pi} (\partial_x \Theta_2)^2] dx, \nonumber \\
&[\Theta_1(x'), \partial_x \Theta_2(x)] = 2\pi i \delta(x'-x), 
\end{align}
and $\mz_2$ symmetry transformation,
\begin{align}
&S^{-1} e^{i \Theta_1} S = e^{i \Theta_2}, \quad \quad S^{-1} e^{i \Theta_2} S = e^{i \Theta_1}
\label{h0101}
\end{align}
(Here, we have chosen the velocity matrix $V$ to be of the form $V = \bpm v & 0 \\ 0 & v \epm$ for simplicity).

We will now show that we can gap the above Hamiltonian $H$ by adding the following $\mz_2$ symmetric perturbation:
\begin{align}
\frac{U}{\sqrt{2}} \int_{-\infty}^{\infty} (\cos(\Theta_1)+\cos(\Theta_2)) dx
\label{symmpert}
\end{align}
Notice that this perturbation is \emph{not} of the null vector type described in section \ref{gappertsect}, so we need to do some additional work to see that it opens up a gap. 

Our argument rests on three claims: (i) $H$ can be gapped by adding the (symmetry-breaking) perturbation $U \cos(\Theta_1)$; (ii) $H$ has a (hidden) $SU(2)_L\times SU(2)_R$ symmetry; and (iii) this $SU(2)_L\times SU(2)_R$ symmetry group contains an element $\mathcal{U}_0$ with the property that
\beq\label{gapping_symmetry}
\mathcal{U}_0^{-1} \cos(\Theta_1) \mathcal{U}_0 = \frac{1}{\sqrt{2}}(\cos(\Theta_1)+\cos(\Theta_2))
\eeq
Given these claims, it follows immediately that the perturbation (\ref{symmpert}) opens up a gap. Furthermore, the resulting ground state must be non-degenerate since it is non-degenerate in the case of the perturbation $U \cos(\Theta_1)$. Putting these two facts together, we conclude that the perturbation (\ref{symmpert}) gaps the edge theory without breaking the $\mz_2$ symmetry either explicitly or spontaneously, {as we wanted to show.}

All that remains is to justify claims (i)-(iii). Claim (i) follows from the fact that $U \cos(\Theta_1)$ belongs to the general class of gapping perturbations discussed in section (\ref{gappertsect}). As for claim (ii), this follows from the well-known equivalence between the above Hamiltonian (\ref{h0101}) and the level-$1$ $SU(2)$ WZW model.\cite{affleck1986} Likewise, claim (iii) follows from the fact that $\bpm e^{i \Theta_1} \\ e^{i \Theta_2} \epm$ transforms as a doublet under the $SU(2)_R$ symmetry. 


\section{Sufficient conditions for when $|\det(K_\pm)| \cdot 2^{|\text{sig}(K_\pm)|}$ is a perfect square} \label{perfectsqapp}
In this appendix, we prove that $|\det(K_\pm)| \cdot 2^{|\text{sig}(K_\pm)|}$ is a perfect square for any SPT edge theory $(K,W,\chi)$ that has $\nu = 0 \pmod{1/2}$.

Our proof is essentially a brute force calculation, using the standard form for $K$ and $W$, given in Eq.~(\ref{k_final_maintext}). The first step is to recall that in this basis, $K_\pm$ are given by
\begin{align*}
K_+ = \bpm C & 2D \\ 2D^T & 2(E+F) \epm, \quad  K_- = \bpm A & 2B \\ 2B^T & 2(E-F) \epm
\end{align*}
Next we define two related matrices:
\begin{align}
J_+ = \bpm C & 2D \\ D^T & (E+F) \epm, \quad J_- = \bpm A & 2B \\ B^T & (E-F) \epm
\end{align}
We note that
\begin{align}
K_\pm = \bpm \mathbf{1_{n_+ - m}} & 0 \\ 0 & 2 \cdot \mathbf{1_{m}} \epm \cdot J_\pm
\end{align}
so $\det(K_\pm) = 2^{m} \det(J_\pm)$. At the same time, it is easy to see that the eigenvalues of $J_\pm$ are in one-to-one correspondence with the eigenvalues of $K$ within the $W = \pm 1$ subspaces. This implies the identity
\begin{align}
\det(K) = \det(J_+) \det(J_-) 
\end{align}
from which it follows that $\det(J_\pm) = \pm 1$. Hence, we conclude that
\begin{align}
|\det(K_+)| = |\det(K_-)| = 2^{m}
\end{align}

It is also clear that
\begin{align}
\text{sig}(K_\pm) = \text{dim}(K_\pm) = n_\pm \pmod{2}
\end{align}
Thus, in order to show that $|\det(K_\pm)| \cdot 2^{|\text{sig}(K_\pm)|}$ is a perfect square, it suffices to show that $m + n_+$ and $m+n_-$ are both even. We will do this with the help of the following identity (see below for proof):
\begin{equation}
4 \nu = m + n_- \pmod{2}
\label{mid}
\end{equation}
The above identity, together with our assumption that $\nu = 0 \pmod{1/2}$, immediately implies that $m + n_-$ is even. Likewise, the fact that $m+n_+$ is even follows from noting that $n_+ = n_- \pmod{2}$ (since $n_+ + n_- = N$ which is an even integer given that $\text{sig}(K) = 0$).
 
All that remains is to establish Eq.~(\ref{mid}). To this end, we note that the general constraint on $\chi_+$ (\ref{chi_plus_constraint}) implies that
\begin{align}
2 \chi_+ = \bpm 0 \\ 0 \\ d \\ d \epm \pmod{2}, \quad \quad d = \text{diag}(E+F)
\end{align} 
It follows that
\begin{align}
(2 \chi_+)^T  K^{-1} (2 \chi_+) = \bpm 0 & 0 & d & d \epm^T K^{-1} \bpm 0 \\ 0 \\ d \\ d \epm \pmod{4}
\label{eq1chiplus}
\end{align}
Next, we use the relationship between $K$ and $K_+$ to rewrite the expression on the right-hand side as 
\begin{align}
\bpm 0 & 0 & d & d \epm^T K^{-1} \bpm 0 \\ 0 \\ d \\ d \epm &= \bpm 0 & 2 d \epm^T K_+^{-1}  \bpm 0 \\ 2 d \epm \nonumber \\
&= 2 \bpm 0 & d \epm^T J_+^{-1}  \bpm 0 \\ d \epm
\label{eq2chiplus}
\end{align}
Now consider $J_+^{-1}$ modulo $2$. Using the fact that $2D \equiv 0 \pmod{2}$, it is easy to see that
\begin{align*}
J_+^{-1} = \bpm C^{-1} & 0 \\ -(E+F)^{-1} D^T C^{-1} & (E+F)^{-1} \epm \pmod{2}
\end{align*}
where all inverses are taken modulo $2$. It follows that
\begin{align}
\bpm 0 & d \epm^T J_+^{-1}  \bpm 0 \\ d \epm &= d^T (E+F)^{-1} d \pmod{2} 
\label{eq3chiplus}
\end{align}
Next we use a general identity, which applies to any symmetric integer matrix $M$ which is invertible modulo $2$:
\begin{align}
\text{diag}(M)^T M^{-1} \text{diag}(M) = \text{dim}(M) \pmod{2}
\label{mod2id}
\end{align}
(One way to prove this to use the fact that any such $M$ can be put into a canonical form $M = \bpm \mathbf{1_k} & 0 & 0 \\ 0 & 0 & \mathbf{1_l} \\ 0 & \mathbf{1_l} & 0 \epm$, after a change of basis). Applying Eq.~(\ref{mod2id}) to $M = E+F$, we conclude that 
\begin{align}
d^T (E+F)^{-1} d = m \pmod{2}
\label{eq4chiplus}
\end{align}
Putting together Eqs.~(\ref{eq1chiplus}-\ref{eq3chiplus}) and (\ref{eq4chiplus}), we derive
\begin{align}
4 \chi_+^T  K^{-1} \chi_+ = 2 m \pmod{4}
\end{align}
Therefore
\begin{align}
4 \nu &= 2 \chi_+^T K^{-1} \chi_+ + \text{sig}(K (1-W)) \nonumber \\
&= m + n_- \pmod{2} 
\end{align}
where we have used the fact that $\text{sig}(K (1-W)) = n_- \pmod{2}$ in the second equality. This completes our proof of Eq.~(\ref{mid}).


\section{General proof that $K_{\pm}$ has null vectors}\label{K_plusminus_null_vectors}

Let $(K,W,\chi)$ be a chiral boson edge theory with $\mz_2$ symmetry. In this appendix, we prove that the two matrices $K_+$ and $K_-$ are guaranteed to have a complete set of null vectors as long as $\text{sig}(K(1-W)) = 0$ and the edge theory satisfies conditions (i)-(iii) of our criterion. 

We will give a separate proof for the case where $(K,W,\chi)$ is a \emph{bosonic} edge theory and the case where it is a \emph{fermionic} edge theory. We will start with the bosonic case, where we will not need condition (3) at all to prove the result. The first step is to note that condition (2) of the criterion implies that there exists a set of integer vectors $\mathcal{L}$ with the following properties:
\begin{enumerate}
\item{$l^T K^{-1} l'$ is an integer for any $l, l' \in \mathcal{L}$.}

\item{$l^T K^{-1} l$ is an \emph{even} integer for any $l \in \mathcal{L}$.}

\item{If $l'$ is not equivalent to any element of $\mathcal{L}$, then $l^T K^{-1} l'$ is non-integer for some $l \in \mathcal{L}$.}

\item{If $l \in \mathcal{L}$, then $W^T l$ is equivalent to some $l' \in \mathcal{L}$.} 
\end{enumerate}
Define a set $\Gamma$ by
\begin{equation}
\Gamma = \{l + K\Lambda: \ l \in \mathcal{L}, \ \Lambda \in \mathbb{Z}^{N}\}
\label{gammapmdef}
\end{equation}
Also, define two sets $\Gamma_\pm$ by 
\begin{equation}
\Gamma_\pm = \{v : \ v \in \Gamma, \ W^T v = \pm v\}
\end{equation}
These sets have several important properties. First, all three are integer lattices, with $\Gamma$ being of dimension $N$ and $\Gamma_\pm$ being of dimension $n_\pm$. Second, $W^T \Gamma = \Gamma$, since this follows immediately from property (4) above. Finally, $\Gamma, \Gamma_\pm$ satisfy
\begin{equation}
\Gamma \subset \frac{1}{2}(\Gamma_+ + \Gamma_-)
\label{gammapmprop}
\end{equation}
In other words, every vector $v \in \Gamma$ can be written as a sum $v = \frac{1}{2}(v_+ + v_-)$ for some $v_\pm \in \Gamma_\pm$. This follows immediately by setting $v_\pm = (v \pm W^T v)$.

To proceed further, we note that since $\Gamma_\pm$ are integer lattices of dimension $n_\pm$, they can be represented as $\Gamma_\pm = U_\pm \mathbb{Z}^{n_{\pm}}$ where $U_\pm$ are $N \times n_\pm$ integer matrices. Now define two matrices 
\begin{align}
\tilde{K}_\pm \equiv U_\pm^T K^{-1} U_\pm 
\end{align}
It is easy to see that $\tilde{K}_\pm$ are symmetric integer matrices with vanishing signature and only even numbers on the diagonal. Indeed, the fact that $\tilde{K}_\pm$ are integer matrices follows from property (1) of $\mathcal{L}$. Also, the fact that $\tilde{K}_\pm$ have only even elements on the diagonal follows from property (2) together with the fact that $K$ has only even elements on the diagonal. Finally, the fact that $\tilde{K}_\pm$ have vanishing signature follows from our assumption that $K(1-W)$ and $K$ have vanishing signature.

In addition to the above properties, $\tilde{K}_\pm$ also have the property that $2 \tilde{K}_\pm^{-1}$ are \emph{integer} matrices. We now prove this claim for $\tilde{K}_+$ --- the proof for $\tilde{K}_-$ is similar. To begin, suppose that $x$ is a $n_+$ component vector with the property that $\tilde{K}_+ x$ is an integer vector. Then it follows that $v_+ K^{-1}U_+ x$ is integer for all $v_+ \in \Gamma_+$. At the same time, it is clear that $v_- K^{-1} U_+ x = 0$ for all $v_- \in \Gamma_-$. Combining these two observations we deduce that $v K^{-1} U_+ x$ is a  \emph{half-integer} for all $v \in \Gamma$: this follows from Eq.~(\ref{gammapmprop}) above. The latter property implies that $2 U_+ x \in \Gamma$, as this follows from property 3 above. At the same time, we know that $2 U_+ x \in \Xi_+$, so we conclude that $2 U_+ x \in \Gamma_+$. It then follows that $2x \in \mathbb{Z}^{n_+}$. Thus, we have shown that if $\tilde{K}_+ x$ is an integer vector then $2x \in \mathbb{Z}^{n_+}$. The claim follows immediately.

To summarize: we have shown that $\tilde{K}_\pm$ are symmetric integer matrices with vanishing signature, with only even elements on the diagonal, and with the property that $2\tilde{K}_\pm^{-1}$ are integer matrices. This means that if we think of $\tilde{K}_\pm$ as $K$-matrices, then the corresponding topological phases are bosonic, have vanishing chiral central charge, and have the property that $a \times a = 1$ for all anyons $a$. Therefore, according to Lemma \ref{lemma1}, these topological phases must have Lagrangian subgroups. This is turn means that $\tilde{K}_\pm$ have a complete set of linearly independent null vectors $\{v_\pm^{(1)},...,v_\pm^{(n_\pm/2)}\}$, according to the results of Ref.~\onlinecite{levin2013protected}. Now define 
\begin{equation}
w_{\pm}^{(i)} = \det(K) \cdot K^{-1} U_\pm v_\pm^{(i)}
\end{equation}
By construction, the $w_{\pm}^{(i)}$'s are null vectors for $K$. At the same time, they belong to $\Xi_\pm$, so they can be represented as
\begin{equation}
w_{\pm}^{(i)} = V_\pm \bar{\Lambda}_{\pm}^{(i)}
\end{equation}
We can now see that $\{\bar{\Lambda}_\pm^{(1)},...,\bar{\Lambda}_\pm^{(n_\pm/2)}\}$ are null vectors for $K_\pm$. This completes the proof in the bosonic case.

Next, consider the fermionic case. Following the same arguments as in the bosonic case, it is easy to see that $\tilde{K}_+$ and $\tilde{K}_-$ are symmetric integer matrices with vanishing signature; the only difference is that now both of these matrices have at least one \emph{odd} element on the diagonal. This means that $\tilde{K}_\pm$ describe \emph{fermionic} topological phases with vanishing chiral central charge and with the property that $a \times a = 1$ for all anyons $a$. 

At this point, it is natural to try to use Lemma \ref{lemma3} to conclude that the topological phases corresponding to $\tilde{K}_\pm$ have Lagrangian subgroups. If we could do that, then we would be done since the remainder of the proof would go through just as in the bosonic case. The only catch is that in order to apply Lemma \ref{lemma3}, we need to establish that $|\det(\tilde{K}_\pm)|$ is a perfect square. We now prove that this is the case, using condition (iii) of the criterion.

The first step is to note that since $K \Xi_\pm \subset \Gamma_\pm$, there must exist integer $n_\pm \times n_\pm$ matrices $T_\pm$ such that
$K V_\pm = U_\pm T_\pm$. It then follows that 
\begin{align}
K_\pm = T_\pm^T \tilde{K}_\pm T_\pm
\end{align}
so that 
\begin{align}
|\det(K_\pm)| = |\det(\tilde{K}_\pm)| \cdot |\det(T_\pm)|^2
\end{align}
At the same time, we know that $|\det(K_\pm)|$ is a perfect square by condition (iii) of our criterion (since $\text{sig}(K_\pm) = 0$ by assumption). Combining this with the above identity, we conclude that $|\det(\tilde{K}_\pm)|$ must also be a perfect square. This completes the proof in the fermionic case.

\end{appendix}

\bibliography{thetaPhi-2}

\end{document}